\documentclass[hyper,letterpaper]{JHEP3} 

\JHEPspecialurl{http://jhep.sissa.it/JOURNAL/JHEP3.tar.gz}

\newcommand{\gsim}{ \mathop{}_{\textstyle \sim}^{\textstyle >} }

\newcommand{\be}{\begin{eqnarray}}
\newcommand{\ee}{\end{eqnarray}}

\usepackage{epsfig}

\newcommand\fverb{\setbox\pippobox=\hbox\bgroup\verb}
\newcommand\fverbdo{\egroup\medskip\noindent%
            \fbox{\unhbox\pippobox}\ }
\newcommand\fverbit{\egroup\item[\fbox{\unhbox\pippobox}]}

\newbox\pippobox

\title{Consequences of a Dark Disk for the \textit{Fermi} and \textit{PAMELA} Signals in Theories with a Sommerfeld Enhancement}

\author{Ilias Cholis and Lisa Goodenough\\
         Center for Cosmology and Particle Physics, Dept. of Physics, New York University, \\ 
         New York, NY 10003\\}

\abstract{Much attention has been given to dark matter explanations of the \textit{PAMELA} positron fraction and \textit{Fermi} electronic excesses.  For those theories with a TeV-scale WIMP annihilating through a light force-carrier, the associated Sommerfeld enhancement provides a natural explanation of the large boost factor needed to explain the signals, and the light force-carrier naturally gives rise to hard cosmic ray spectra without excess $\pi^0$-gamma rays or anti-protons.  The Sommerfeld enhancement of the annihilation rate, which at low relative velocities $v_{rel}$ scales as $1/v_{rel}$, relies on the comparatively low velocity dispersion of the dark matter particles in the smooth halo.  Dark matter substructures in which the velocity dispersion is smaller than in the smooth halo have even larger annihilation rates.  N-body simulations containing only dark matter predict the existence of such structures, for example subhalos and caustics, and the effects of these substructures on dark matter indirect detection signals have been studied extensively.  The addition of baryons into cosmological simulations of disk-dominated galaxies gives rise to an additional substructure component, a dark disk.  The disk has a lower velocity dispersion than the spherical halo component by a factor $\sim 6$, so the contributions to dark matter signals from the disk can be more significant in Sommerfeld models than for WIMPs without such low-velocity ehancements. We consider the consequences of a dark disk on the observed signals of $e^{+}e^{-}$, $p \bar{p}$ and $\gamma$-rays as measured by \textit{Fermi} and \textit{PAMELA} in models where the WIMP annihilations are into a light boson.  We find that both the \textit{PAMELA} and \textit{Fermi} results are easily accomodated by scenarios in which a disk signal is included with the standard spherical halo signal.  If contributions from the dark disk are important, limits from extrapolations to the center of the galaxy contain significant uncertainties beyond those from the spherical halo profile alone.}

\preprint{}

\begin{document}


\section{Introduction}\label{sec:intro}
Much excitement has been generated over the past year or so by the release of results from several cosmic ray experiments. \textit{PAMELA} \cite{Adriani:2008zr, PAMELA2, PAMELA, Adriani:2010ib} has shown a sharp upturn in the positron fraction above 10 GeV.  \textit{Fermi} \cite{Abdo:2009zk} has published results indicating an excess in the total electronic spectrum $e^+ + e^-$ in the 100 GeV to 1 TeV range, while HESS has shown that the spectrum falls off sharply above 1 TeV \cite{Collaboration:2008aaa, Aharonian:2009ah}. Thermal WIMPs would naturally predict such features in the electron and positron measurements.  However, conventional WIMPs annihilate far too little to yield these large signals, which require boost factors of $O(100)$ above the thermal cross-section of $3 \times 10^{-26}\; \rm cm^{3} s^{-1}$ \cite{Bergstrom:2008gr, Cholis:2008hb, Cirelli:2008jk, Barger:2009yt, Cline:2010ag, Cirelli:2010nh}.  Moreover, the absence of an excess in antiprotons up to 100 GeV in the \textit{PAMELA} \cite{Adriani:2008zq} data essentially excludes models of dark matter which would produce the $e^{+}e^{-}$ excess through annihilations to standard model modes, since such models also have significant hadronic branching fractions \cite{Cirelli:2008pk, Donato:2008jk}.

An appealing possibility is for the dark matter (DM) to annihilate into a new, light ($m \lesssim 1$ GeV) boson \cite{Cholis:2008vb}.  This would result in a hard lepton spectrum and would kinematically prevent or suppress $\bar p$ production.  Such annihilation modes easily fit the \textit{PAMELA} and \textit{Fermi} electron results \cite{Cholis:2008wq, Cholis:2008vb}, while also producing an interesting synchrotron ``microwave haze'' signal, which has been observed by \textit{WMAP} \cite{Finkbeiner:2003im, Finkbeiner:2004us, Dobler:2007wv}, and additionally yielding a significant $\gamma$-ray signal \cite{Dobler:2009xz} from inverse Compton and final state radiation that may be observable at \textit{Fermi}.  A key consequence of including a new force in the dark sector is that at low velocities an annihilation rate much larger than that expected for a thermal WIMP can result from the Sommerfeld enhancement \cite{ArkaniHamed:2008qn, Lattanzi:2008qa, MarchRussell:2008yu} or from capture into bound ``WIMPonium'' states \cite{Hisano:2004ds, Cirelli:2007xd, Pospelov:2008jd, MarchRussell:2008tu}.  Sommerfeld-enhanced annihilation rates increase like $1/v_{rel}$ at small $v_{rel}$, until the non-zero mass of the mediator $\phi$ cuts off the enhancement.  As argued in \cite{ArkaniHamed:2008qn}, an important feature of such low-velocity enhancements is the added significance of substructure.  Because the velocity dispersion in the dark disk can be significantly lower than in the smooth halo, up to $\sim 6$ times smaller, the annihilation cross-section and, therefore, annihilation rate, can be up to $\sim 6$ larger when the enhancement scales as $1/v_{rel}$.  Add to this effect the increase in the annihilation rate due to the fact that the DM density in substructure is larger than in the smooth halo, and the role of substructure can indeed be substantial.

The spectra of positrons and electrons can be greatly affected by the presence of substructure.  Because distant ($d \gsim 1$ kpc) positrons and electrons tend to lose their energy before reaching Earth, the high energy electronic signals, such as those measured by \textit{PAMELA} and \textit{Fermi}, are dominated by nearby sources. If there is nearby substructure, the observed spectra can be modified \cite{Hooper:2008kv}. Quantifying the effects, even for standard WIMPs, is a great challenge; the {\em Via Lactea} N-body simulation \cite{Diemand:2008in, Kuhlen:2008qj, Diemand:2006ik} argues for a 1\% chance of a subhalo giving a boost of 10 for conventional WIMPs, while the {\em Aquarius} simulation sees lower boosts \cite{Springel:2008cc}.  For WIMP scenarios with a Sommerfeld enhancement, quantifying the overall effect is more difficult.  The Sommerfeld enhancement saturates at some velocity, below which the $1/v$ behavior is no longer valid.  If the saturation occurs at $v \sim$ 200 km/s, as suggested in \cite{Slatyer:2009yq}, then the annihilation cross-section is the same in the smooth DM halo, which has a velocity dispersion of $\sim 220$ km/s, as in substructure with lower velocity dispersion.  If the saturation occurs at much lower velocities, the annihilation cross-section in substructure with the lowest velocity dispersion can be substantially larger than in the smooth halo.  Since it is the smallest halos that have the lowest velocity dispersion, the smallest substructure can play a dominant role.  This results in significant sensitivity to difficult-to-resolve small scales, and additionally to the unknown mass of the mediator $\phi$, which cuts off the enhancement, and cutoffs in the primordial power spectrum.  Other complications, such as resonances in the annihilation cross-section for specific parameters, can make quantifying the effects nearly impossible.

Nonetheless, it is important to attempt to understand what effects substructure can have on the present signals, in the context of Sommerfeld-enhanced annihilation channels. In this paper, we consider the effects of the addition of a ``dark disk'', as proposed by \cite{Read:2008fh, Read:2009iv}, to the smooth spherical halo (see also \cite{Pato:2010yq}).  Cosmological simulations that model only the dark matter particles are neglecting the baryons that dominate the mass of the galaxy interior to our location at the Solar Circle at $R_{\odot}=8.5$ kpc.  The dominant mass component is the stellar disk, and it plays an important role in the accretion history of the galaxy.   \cite{Read:2008fh, Read:2009iv} have shown that dynamical friction between the galactic disk and satellite galaxies causes the satellites to be preferentially dragged into the plane of the disk.  Tidal forces subsequently disrupt the satellites, and the baryons and dark matter are accreted into disk-like structures.  The resulting dark disk, which is kinematically similar to the stellar thick disk, has a velocity dispersion of $\sim 30-90$ km/s. 

\begin{figure}[t]
\centering
\begin{tabular}{cc}
\epsfig{figure=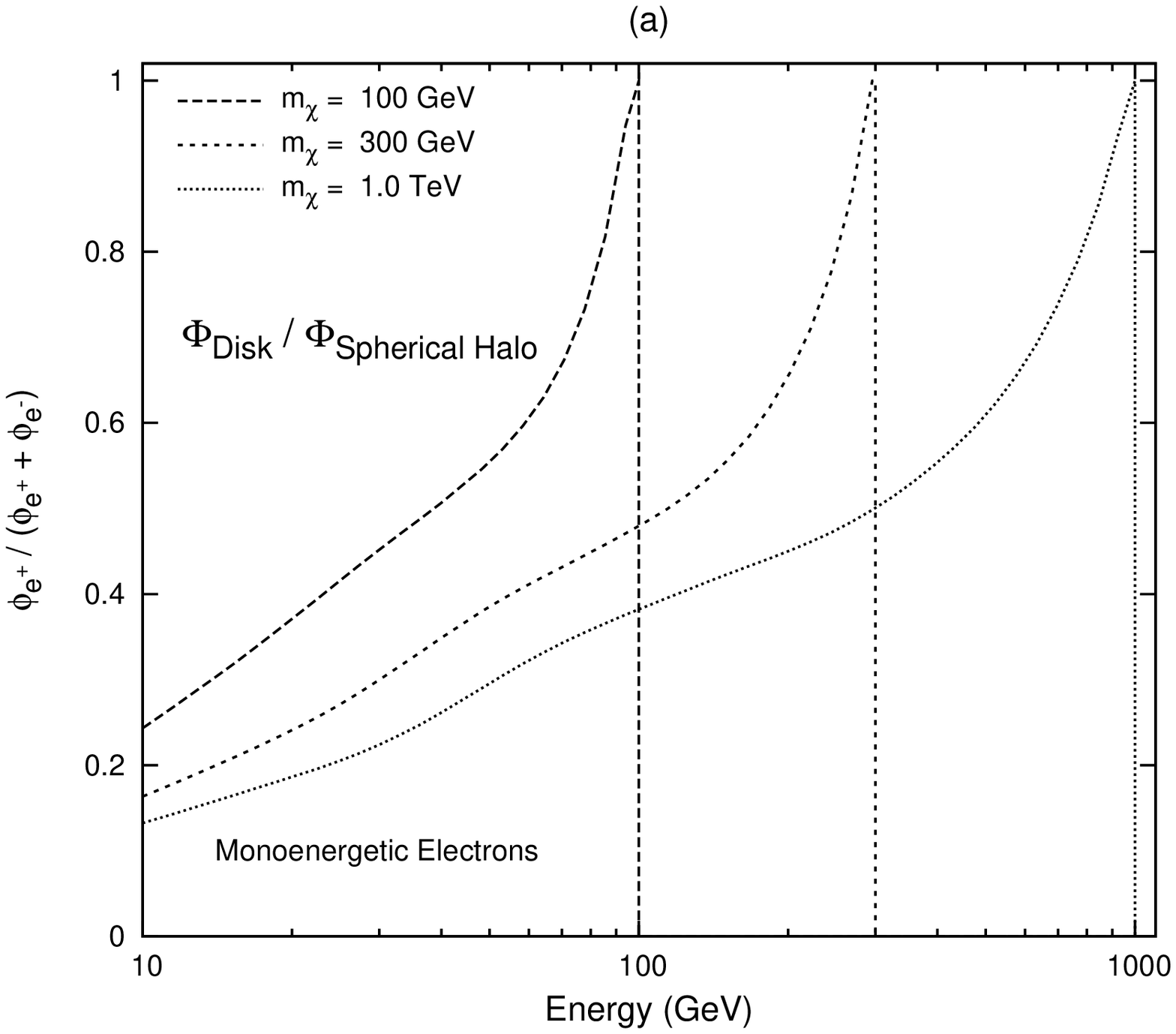,width=3.0in} &
\epsfig{figure=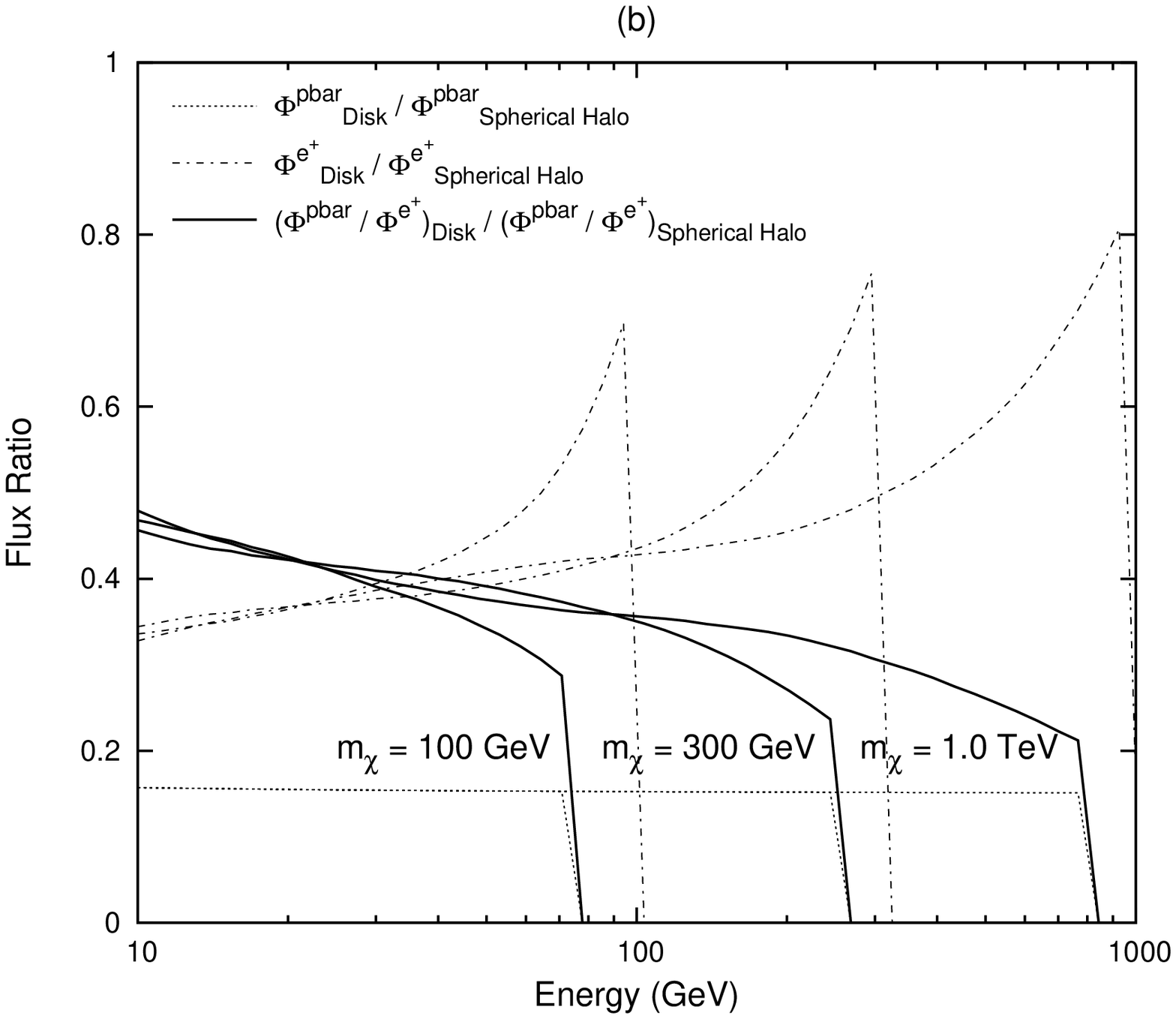,width=3.0in} 
\end{tabular}
\caption{Left: The ratio of the local flux of positrons from DM annihilation for the Dark Disk profile to that for the spherical Einasto profile.  Here we assume monoenergetic electrons from the annihilation $\chi \chi \rightarrow e^{+} e^{-}$, with $m_{\chi}=100\; \rm GeV$, $m_{\chi}=300\; \rm GeV$, and $m_{\chi}=1.0\; \rm TeV$.  All ratios have been normalized to 1 at the highest energy.  Right: The ratio of the local flux of positrons (\textit{dot-dashed}) and antiprotons (\textit{dotted}) from DM annihilation for the Dark Disk profile to that for the spherical Einasto profile for the annihilation channel $\chi \chi \rightarrow W^{+} W^{-}$, with $m_{\chi}=100\; \rm GeV$, $m_{\chi}=300\; \rm GeV$, and $m_{\chi}=1.0\; \rm TeV$.  Also, we show the ratio of ratios $(\Phi_{\bar p}/\Phi_{e^+})_{Disk}/(\Phi_{\bar p}/\Phi_{e^+})_{Einasto}$ (\textit{solid}).}
\label{fig:DiskHaloratio}
\end{figure}

The presence of a dark disk in our galaxy can have consequences for the observed cosmic ray (CR) spectra.\footnote{For a discussion of the effects of a dark disk on neutrino signals, see \cite{Bruch:2009rp}.}  The disk does not have the relatively large increase in mass density at the Galactic Center that the spherical halo has, and it extends only $\sim 1-2$ kpc from the Galactic Plane, so it sources fewer distant dark matter annihilation products.  As a result, the disk gives a spectral hardening (as viewed here on Earth) of electrons and positrons, which are subject to large energy losses during propagation.  See Fig.~\ref{fig:DiskHaloratio}a in which we plot the ratio of the dark matter positron flux due to annihilations from a pure disk to that from annihilations in a pure spherical halo, $\Phi_{Disk}/\Phi_{Einasto}$.  The ratio increases with energy, indicating that the relative contribution to the highest energy positrons is larger for the disk, which has a larger relative dark matter density in the Galactic Plane than a spherical halo.  

The spectral hardening of the electrons and positrons is not accompanied by a corresponding hardening of the proton and antiprotons, as the latter are not subject to the same large energy losses during propagation.  In this way, substructure can change the relative numbers of annihilation products of a given energy reaching our detectors on Earth.  See Fig.~\ref{fig:DiskHaloratio}b. The ratio of high energy antiprotons to positrons for the Einasto profile is $\sim 1/3$ that for a dark disk.  This can be understood in the following way.  Compared to positrons of the same energy, high energy anti-protons have energy losses that are smaller by a factor of $(\frac{m_{e}}{m_{p}})^2$, while their diffusion lengths are equal in the ultra-relativistic limit.\footnote{Within the validity of approximating  the Klein-Nishina cross-section $\sigma_{K-N}$ with the Tompson cross-section $\sigma_{T}$, $\sigma_{KN} \approx \sigma_{T}$.} Within an infinitely large and homogeneous diffusion zone, antiprotons diffuse a distance $\frac{m_{p}}{m_{e}}$ (on average) larger than that of positrons with the same injected and observed energies. Models of our Galaxy indicate a diffusion zone of half-width $L$ between $\sim 2$ and $\sim 10$ kpc.  Protons and antiprotons of $E\sim 100$ GeV measured by \textit{PAMELA} propagate to us from the entire volume of the assumed diffusion zone.  On the other hand, electrons and positrons with energies $\sim 1$ TeV measured by \textit{Fermi} originate from a region within $\sim 1-2$ kpc of Earth, assuming their injection energies are not significantly higher. Thus, if a significant component of DM annihilation comes from a dark disk profile with a width of $\sim 1-2$ kpc, the observed antiproton flux will be suppressed relative to the predicted flux from a spherical halo alone, while the observed positron flux, being significantly more ``local'' in origin, will not deviate greatly from that predicted in the spherical halo scenario.

The flux of photons coming from dark matter is highly dependent on the spatial distribution of the dark matter in the Galaxy.  As an example, consider the differences between the fluxes expected from dark matter annihilation as compared to dark matter decay.  Since dark disk profiles have a much lower density at the center of the Galaxy than Einasto spherical profiles \cite{Einasto} (see Fig.~\ref{fig:SHM_DD}), a dark disk that gives an $\mathcal{O}(1)$ contribution to the local $e^{\pm}$ fluxes at $E > 500$ GeV, contributes significantly less to $\gamma$-rays and synchrotron radiation in regions within the inner 2 kpc.  Thus, using local $e^{\pm}$ fluxes to find DM annihilation rates results in smaller fluxes of $\gamma$-rays and synchrotron radiation from the center of the Galaxy for dark disks.  Because of this, a significant dark disk component in models with a Sommerfeld enhancement alleviates some the constraints coming from gamma ray measurements of the Galactic Center and Galactic Ridge regions.  As a consequence, while the Galactic Center places the tightest constraints on the annihilation rate for cusped, spherical profiles, the Galactic high latitudes
(specifically, the isotropic diffuse flux there) usually places the tightest constraints on the annihilation rate for the disk profiles.

\section{Definitions and Parameters}\label{sec:defns}

\subsection{Dark Matter Halo Models}

In the $\Lambda$CDM cosmological simulations of \cite{Read:2008fh}, the formation of the stellar thick disk through accretion can be accompanied by the formation of a dark matter disk.  The fits of the dark disk (DD) component can be described by an exponential function and parametrized by half mass scale lengths.  We consider a disk component based on the fits of \cite{Read:2008fh}, and assume a cylindrically symmetric dark disk density profile of the form

\begin{equation}
\label{eq:DD_eq}
	\rho(R,z)=\rho_{0} \exp\left[\frac{1.68 \left(R_{\odot}-R\right)}{R_{1/2}}\right]\exp\left[-\frac{0.693\left|z\right|}{z_{1/2}}\right],
\end{equation}
where $R_{1/2}=11.7$ kpc and $z_{1/2}=1.5$ kpc are the half mass scale lengths in the Galactic Plane and perpendicular to the Galactic Plane, respectively, and $R_{\odot}=8.5$ kpc.  Here $R$ is the cylindrical radial coordinate.

\begin{figure}[t]
\centering
\begin{tabular}{cc}
\epsfig{figure=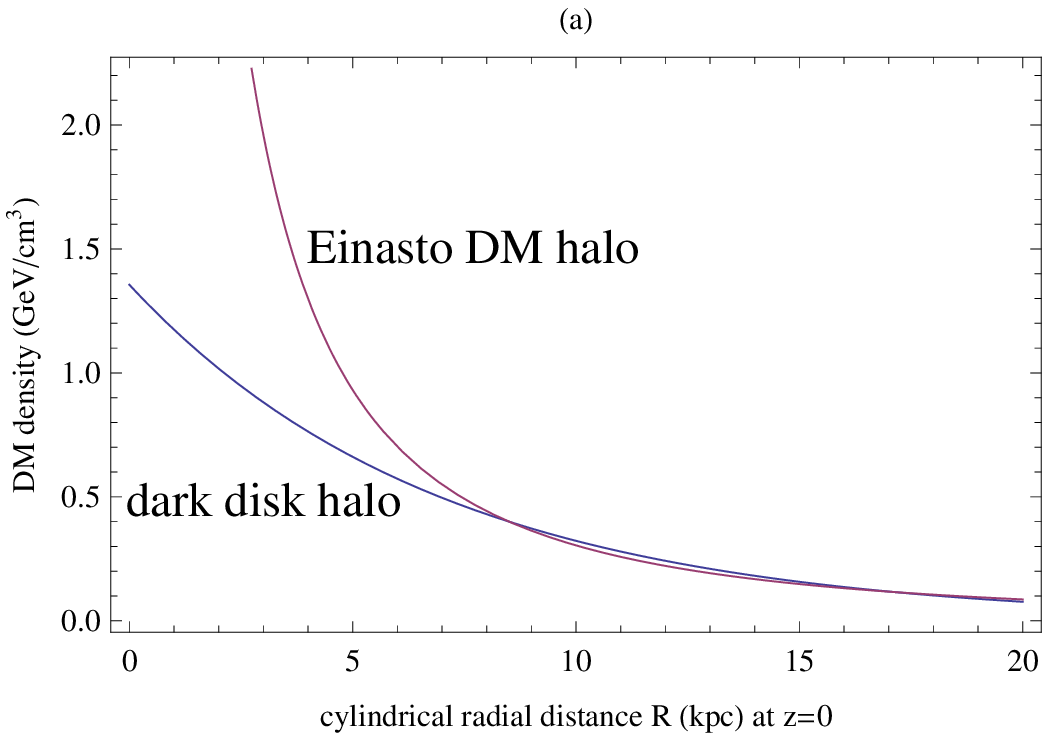,width=3.0in} &
\epsfig{figure=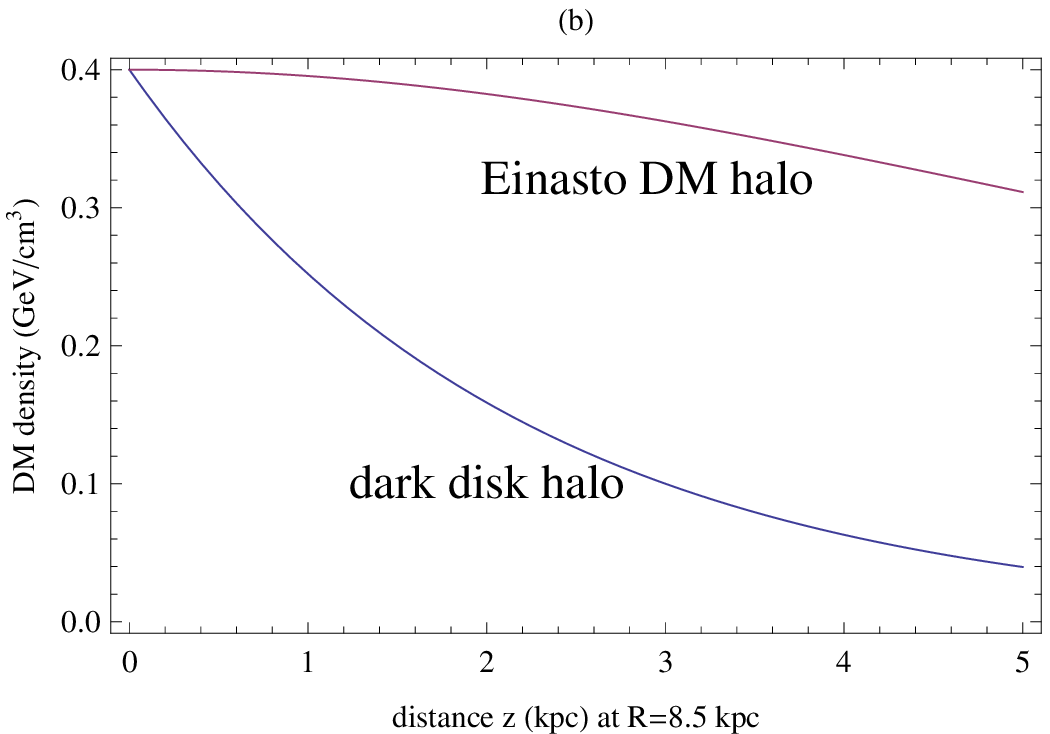,width=3.0in} 
\end{tabular}
\caption{Left: Radial dependence of DM density for the Einasto profile and the Dark Disk profile at $z=0$. Right: $z$ dependence of DM density for the Einasto profile and the Dark Disk profile at $r_{cylindrical}=R_{\odot}$.}
\label{fig:SHM_DD}
\end{figure}

For the spherically symmetric halo model (SH) we use the Einasto profile \cite{Einasto} following Merritt et al. \cite{Merritt:2005xc}

\begin{equation}
\label{eq:SHM_eq}
	\rho(r)=\rho_{0} \exp\left[-\frac{2}{\alpha}\left(\frac{r^{\alpha}-R_{\odot}^{\alpha}}{r_{-2}^{\alpha}}\right)\right],
\end{equation}
with $\alpha = 0.17$ and $r_{-2}=25 \; \rm kpc$.  $r$ is the spherical radial distance from the Galactic Center (GC).
We take the value of the local density for both the spherical and disk components to be $\rho_{0}=0.4 \; \rm GeV cm^{-3}$~\cite{Catena:2009mf}. (See discussion below.)
In Fig.~\ref{fig:SHM_DD} we show the dependence of these DM profiles on $R$ for $z=0$, i.e. in the Galactic Plane, and on $z$ for $R=8.5$ kpc, i.e. perpendicular to the Galactic Plane at our location.

According to \cite{Read:2008fh}, values for the ratio of the local DM density in the dark disk to the local DM density in the spherical halo $\rho_{0_{DD}}/\rho_{0_{SH}}$ range from 0.2 up to more than 1.5. The highest values of $\rho_{0_{DD}}/\rho_{0_{SH}}$ occur in simulations when the thick stellar disk has a higher mass density than the thin stellar disk.  This can result from an increase in the mass included in the stars of the thick disk, or from a thin disk heated by very massive, high-redshift mergers, which results in the thick disk getting populated by thin disk stars.  Another, though less likely, cause is the occurrence of multiple prograte and low inclination mergers \cite{Read:2008fh}.

Constraints have been placed on the total local dark matter density from estimations of the local dynamical mass density based on measurements of proper motions and parallaxes of stars. Using the Hipparcos data, \cite{Holmberg:1998xu} derived the local dynamical mass density to be $0.102 \pm 0.010 \; \rm M_{\odot} pc^{-3}$ (95$\%$ C.L.).  Their estimate for the local density of the visible matter is $0.094 \pm 0.017 \; \rm M_{\odot} pc^{-3}$ (67$\%$ C.L.).  Assuming the mean values of these densities, we get $0.008 \; \rm M_{\odot }pc^{-3}$, or equivalently $0.3 \; \rm GeV cm^{-3}$, for the local dark matter density with an error that is larger than the mean value, thus allowing for a dark disk with $\rho_{0_{DD}} \sim \rho_{0_{SH}}$, where $\rho_{0_{SH}} = 0.4 \; \rm GeV cm^{-3}$.  \cite{Creze:1997rj} derived a smaller local dynamical mass density of $0.076 \pm 0.015 \; \rm M_{\odot} pc^{-3}$ using the Hipparcos data, but values of $0.3-0.4 \; \rm GeV cm^{-3}$ are allowed within their uncertainties.  Later analysis of the Hipparcos data by \cite{Holmberg:2004fj} and \cite{Korchagin:2003yk} agree with the constraints imposed by \cite{Holmberg:1998xu}. Recently \cite{Catena:2009mf} suggested a value of $0.4 \; \rm GeV cm^{-3}$ as the best estimate of the local DM density. Also \cite{Salucci:2010qr} calculated the value of the local DM density to be $0.43 \pm 0.11 \pm 0.10 \; \rm GeV cm^{-3}$, while \cite{Weber:2009pt} set an upper limit of $1 \; \rm GeV cm^{-3}$ for the local DM density.

\subsection{Cosmic Ray Propagation}

We use GALPROP, developed by Moskalenko and Strong \cite{Strong:1999sv}, for the propagation of all cosmic ray particle species.  For our calculation of the Inverse Compton Scattering of electrons and positron we use the Interstellar Radiation Field model of \cite{Porter:2005qx}.  For consistency in our comparisons of the spectra from a dark disk and from a spherical halo, we assume the same energy losses, diffusion coefficient, and diffusion zone\footnote{Refer to \cite{Cholis:2008vb} or \cite{Strong:2007nh} for the details of the diffusion parameters.}.  Our choices for the background models and propagation parameters give local CR spectra that agree with local cosmic ray measurements, including the proton, He, C, Fe, B/C, $\rm Be^{10}/Be^{9}$, and sub-Fe/Fe spectra.  Additionally, for the annihilation channels with no DM contribution to antiprotons our background antiproton-over-proton ratio $\bar{p}/p$ agrees with the \textit{PAMELA} $\bar{p}/p$ data.  See Fig.~\ref{fig:WW_antiprotons}a.

In placing constraints on $\bar{p}$ fluxes from DM, we use a background antiproton ratio that is $15\%$ smaller.  The smaller $\bar{p}/p$ background ratio shown in Fig.~\ref{fig:WW_antiprotons}b is consistent with the current constraints from all CR data.  For example, a reduction in the background  $\bar{p}/p$ could be based on uncertainties in the interstellar medium (ISM) distribution of Hydrogen and nuclei, since antiprotons are produced by collisions of CR protons and nuclei with the ISM protons and nuclei.

\subsection{Annihilation Channels}\label{subsec:annihchannels}

We consider two classes of annihilation scenarios, the ``eXciting Dark Matter" (XDM) scenario \cite{ArkaniHamed:2008qn, Cholis:2008qq, Finkbeiner:2007kk} in which annihilation proceeds through a single mediator $\phi$, and annihilation proceeding through two mediators $\Phi$ and $\phi$.  Additionally we consider annihilation directly into muons, $\chi \chi \rightarrow \mu^{+} \mu^{-}$, and annihilation into $W^{+}W^{-}$, which then decay.  We discuss the implications of the $W^{+}W^{-}$ channel for the \textit{PAMELA} antiproton data.

In the XDM scenario, we present results for annihilation\\
\indent{i) into $e^{+}e^{-}$ through a scalar (or vector) $\phi$ with mass $2m_{e}<m_{\phi}<2m_{\mu}$ and a branching ratio (BR) of 0.9\footnote{We expect a $\sim10\%$ decay of scalar  $\phi\rightarrow$ $2\gamma$ due to 1-loop diagrams.} (or 1),}\\
\indent{ii) into $\mu^{+}\mu^{-}$ through a scalar (or vector) $\phi$ with mass $2m_{\mu}<m_{\phi}<2m_{\pi}$ and BR=1,}\\
\indent{iii) into $\pi^{+}\pi^{-}$ through a vector $\phi$ with mass $m_{\phi}=750$ MeV (the form factor for $\phi \rightarrow \pi^{+} + \pi^{-}$ peaks at $m_{\phi} \simeq 750$ MeV \cite{Eidelman:2002kg}),}\\
\indent{iv) and into $e^{+}e^{-}$, $\mu^{+}\mu^{-}$ and  $\pi^{+}\pi^{-}$ with relative BR's 1:1:2, through a vector $\phi$ with mass $m_{\phi}\simeq 650\; \rm MeV$}.

For DM annihilation with subsequent cascade decay through two mediators, we show the case $\chi \chi \rightarrow \Phi \Phi$, with subsequent decays $\Phi \rightarrow \phi \phi$ and $\phi \rightarrow \mu^{+}\mu^{-}$, where $2m_{\mu}<m_{\phi}<2m_{\pi}$ and $m_{\Phi} \sim 10m_{\phi}$, and additionally the similar cascade decay with  $\phi \rightarrow  e^{+}e^{-}$. The spectra of the final $e^{+}e^{-}$ injected into the ISM from these cascade decays are given in Appendix A.

\subsection{Annihilation Rate and Boost Factor}

In the absence of the Sommerfeld enhancement, the total dark matter annihilation rate $\Gamma_{ann}$ from the spherical halo and dark disk is given by (for Majorana fermions)
\begin{equation}
	\Gamma_{ann}=\frac{1}{2}\left(\frac{\rho_{SH}+\rho_{DD}}{m_{\chi}}\right)^{2}\langle\sigma_{ann}\vert v\vert\rangle.
\label{eq:annih_rate_1}
\end{equation}
If the Sommerfeld enhancement is present, Eq.~\ref{eq:annih_rate_1} is modified to
\begin{equation}
	\Gamma_{ann}=\frac{1}{2}\left(\frac{\rho_{SH}}{m_{\chi}}\right)^{2}\langle\sigma_{ann}\vert v\vert\rangle_{SH} + \frac{1}{2}\left(\frac{\rho_{DD}}{m_{\chi}}\right)^{2}\langle\sigma_{ann}\vert v\vert\rangle_{DD} + \left(\frac{\rho_{SH}\cdot \rho_{DD}}{m_{\chi}^{2}}\right)\langle\sigma_{ann}\vert v\vert\rangle_{mixed}.
\label{eq:annih_rate_Somm}
\end{equation}
$\langle\sigma_{ann}\vert v\vert\rangle_{SH}$ and $\langle\sigma_{ann}\vert v\vert\rangle_{DD}$ are in general different, because the velocity dispersion of dark matter is different in the spherical and disk components of the halo.\footnote{For the mixed term ($\propto \rho_{SH}\cdot \rho_{DD}$), we take $\langle\sigma\vert v\vert\rangle_{mixed} = \langle\sigma_{ann}\vert v\vert\rangle_{SH}$.}

We define the boost factor (BF) as the ratio of the thermally averaged annihilation cross-section needed to fit a set of data, $\langle\sigma_{ann}\vert v\vert\rangle_{fit}$, to $3\times 10^{-26}\; \rm cm^{3} s^{-1}$, the expected thermally averaged annihilation cross-section for a WIMP with mass $m_{\chi}\sim 500\; \rm GeV$,

\begin{equation}
	BF=\frac{\langle\sigma_{ann}\vert v\vert\rangle_{fit}}{3\times 10^{-26}\;  \rm cm^{3} s^{-1}}.
\label{eq:boost_eq}
\end{equation}

\section{Results}\label{sec:results}

\subsection{Positrons and Electrons}\label{subsec:electronicsignals}

\vspace{\baselineskip}
\noindent{\textbf{Sommerfeld Enhanced Annihilation Channels}}

\begin{figure}[h!]
\centering
\begin{tabular}{cc}
\epsfig{figure=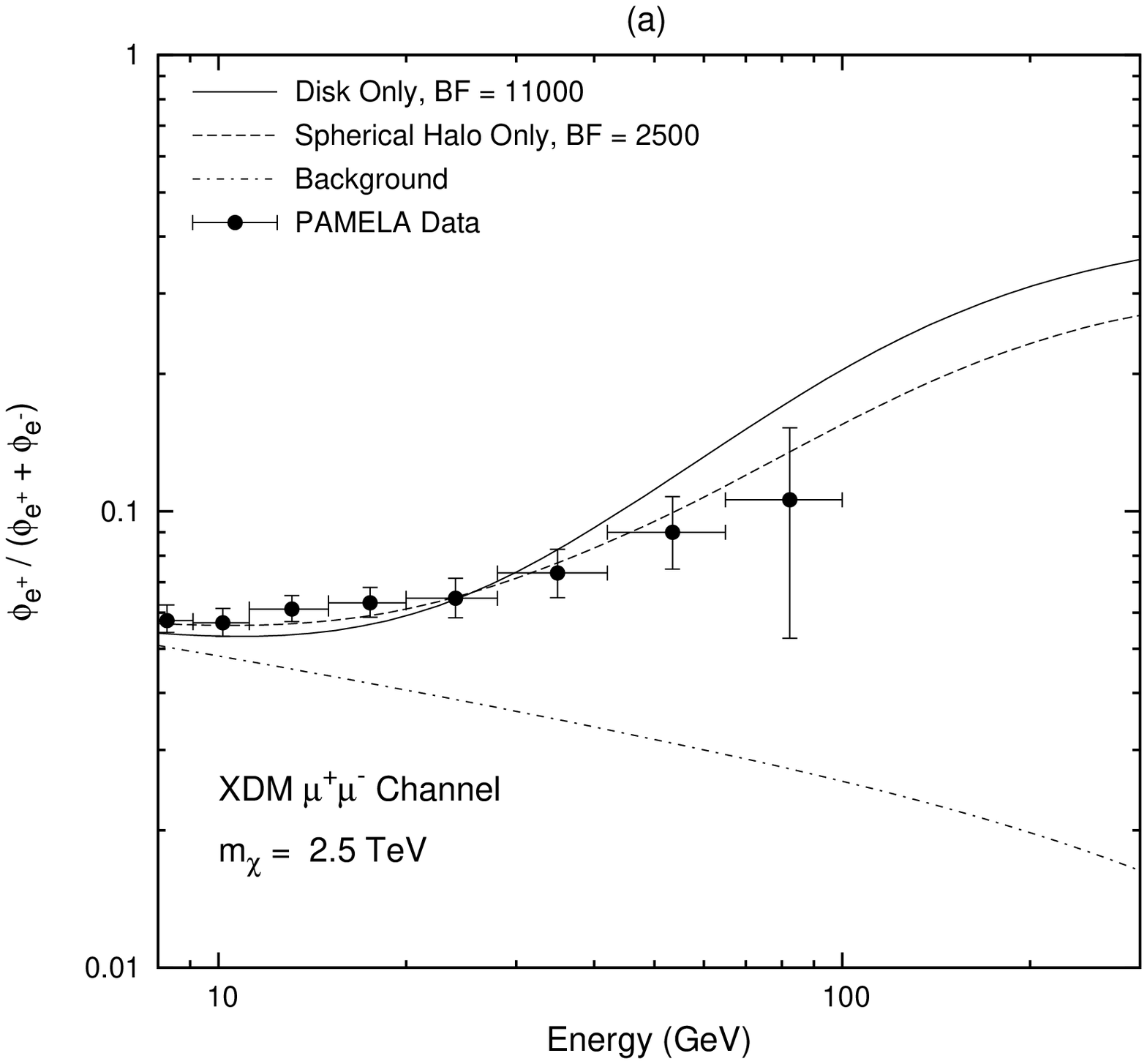,width=3.0in} &
\epsfig{figure=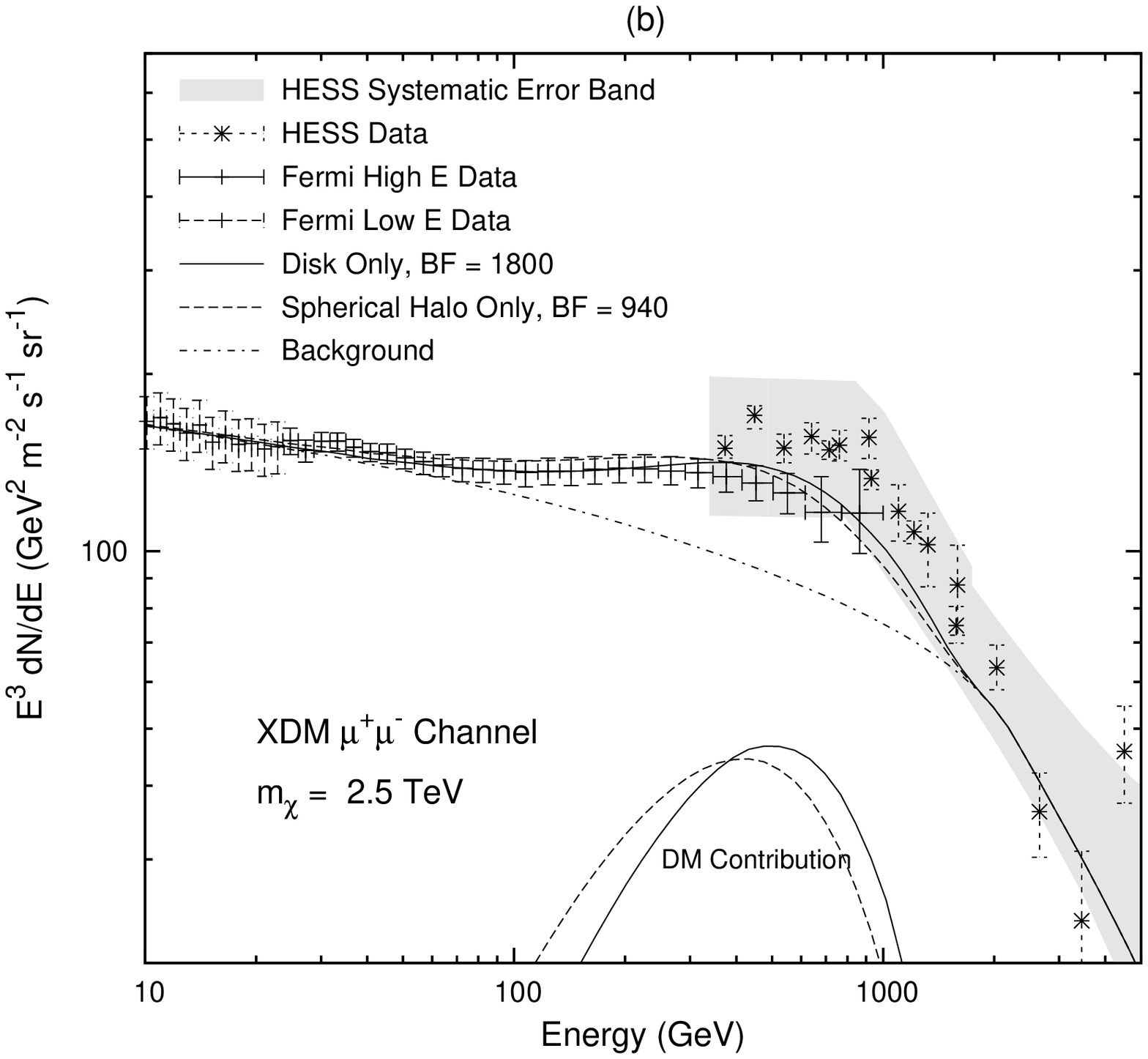,width=3.0in} \\
\vspace{5pt}
\epsfig{figure=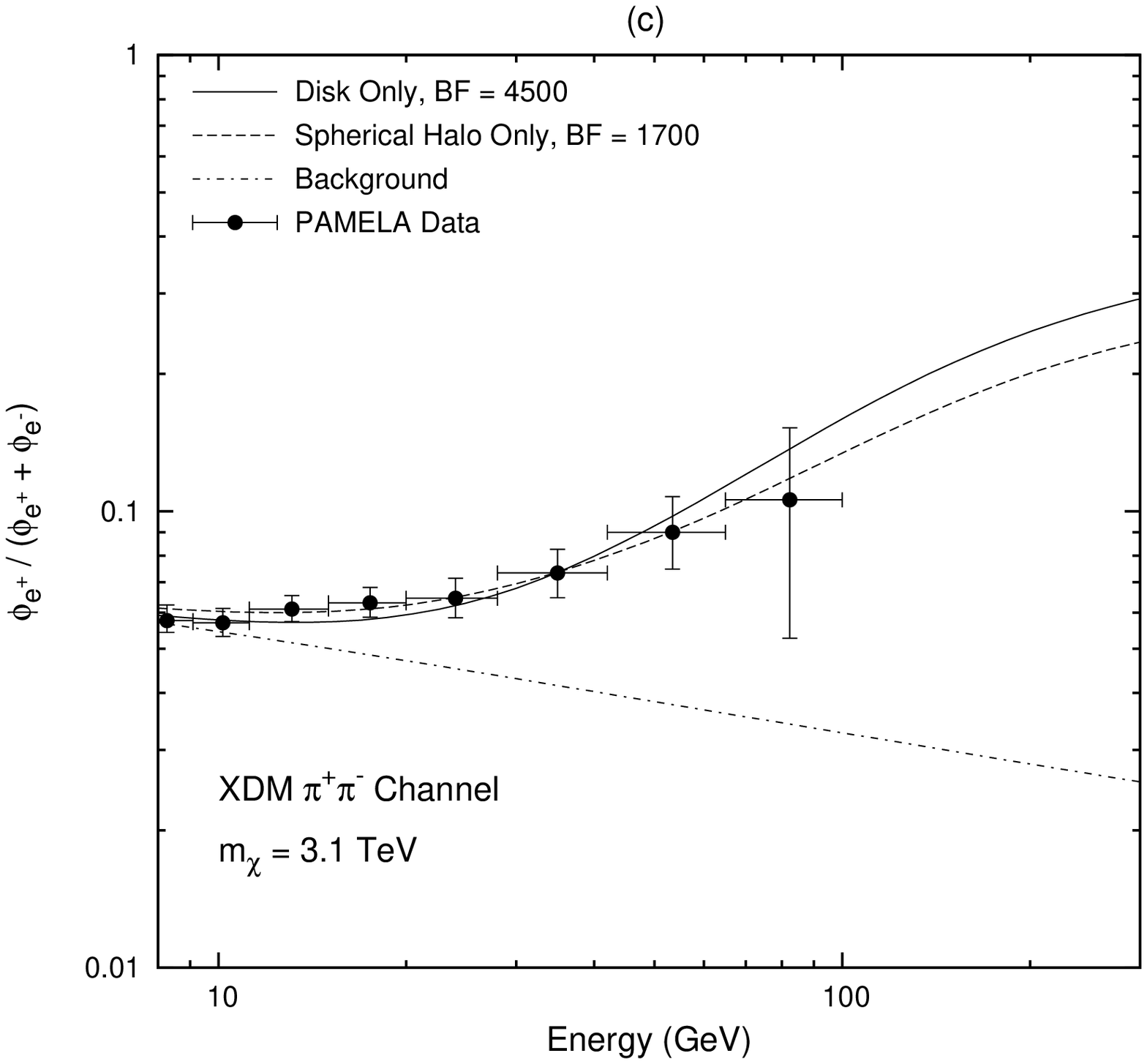,width=3.0in} &
\epsfig{figure=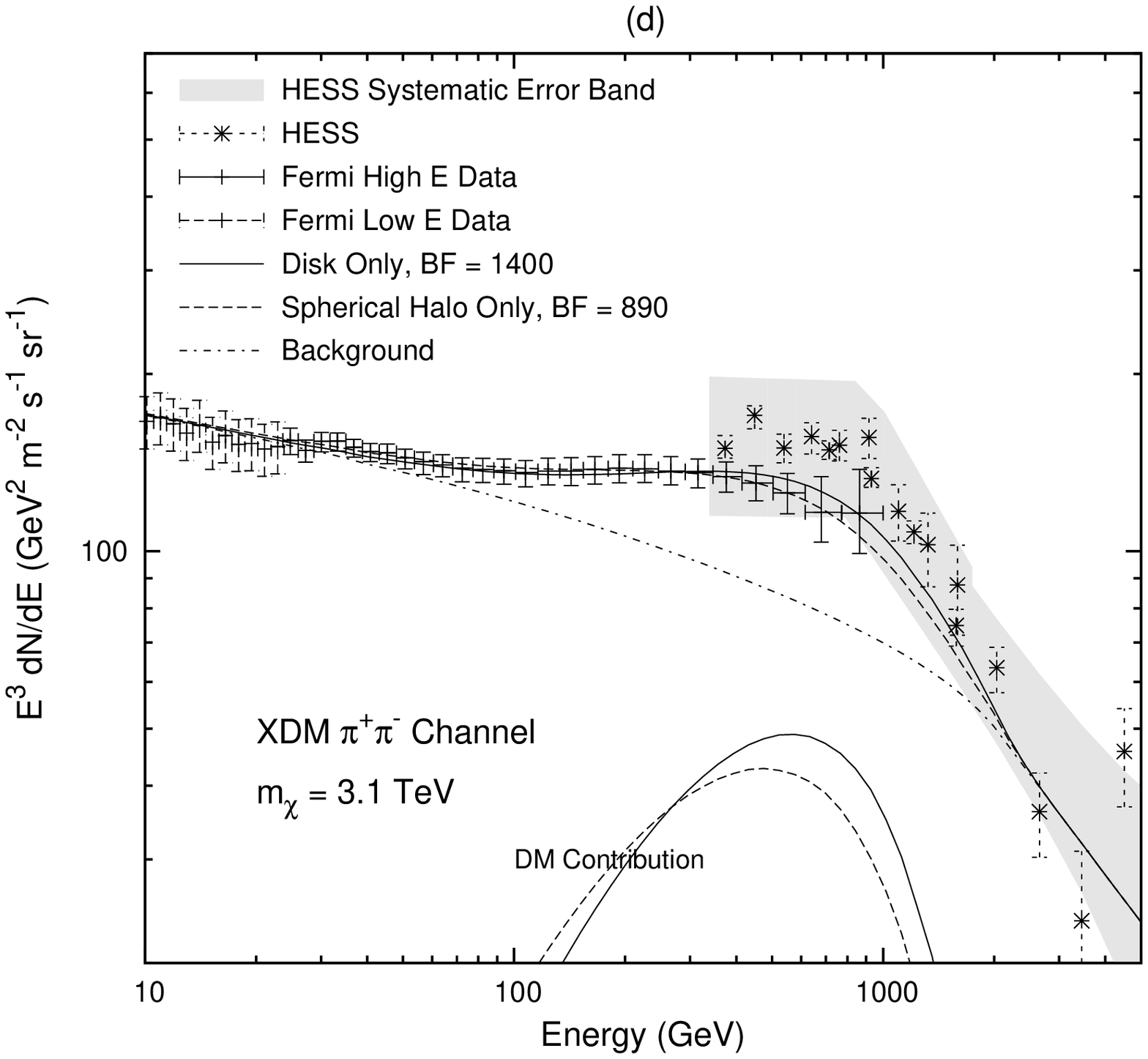,width=3.0in} \\
\end{tabular}
\caption{Positron fraction $\Phi_{e^+}/(\Phi_{e^+}+\Phi_{e^-})$ (left) and total electronic flux weighted by $E^3\; \rm (GeV^2 m^{-2} s^{-1} sr^{-1})$ (right), both as a function of energy.  Results are shown for the spherical Einasto profile (\textit{dashed}) and for a dark disk (\textit{solid}).  Backgrounds are shown in \textit{dot-dashed}.  Top: XDM $\mu^{+} \mu^{-}$ channel with $m_{\chi}=2.5$ TeV.  Bottom: XDM $\pi^{+} \pi^{-}$ channel with $m_{\chi}=3.1$ TeV. \textit{PAMELA} postron fractio data are from \cite{Adriani:2010ib}, while the total electronic flux measured by \textit{Fermi} given in \cite{Abdo:2009zk, FermiLOW}.}
\label{fig:XDM_singlemodes}
\end{figure}
 
As mentioned earlier, the effect of a dark disk on the electronic signal is to produce a harder spectrum at high energies relative to that produced by the spherical Einasto profile.  See Fig.~\ref{fig:DiskHaloratio}a.  This is a generic result for the dark disk profiles of \cite{Read:2008fh} and any spherical halo profile, and it arises because lower energy electrons and positrons, which propagate to Earth over larger distances on average, are probes of regions closer to the Galactic Center where the spherical halo profiles peak significantly more than the disk profiles.  See Fig.~\ref{fig:SHM_DD}a.  This also explains why the boost factor needed to get equal fluxes of dark matter electrons and positrons at energies much less than $m_{\chi}$ is lower for the Einasto profile than for the dark disk, i.e. why the boost factors needed to fit \textit{PAMELA} are lower for the Einasto profile than for the dark disk.
 
\begin{figure}[h!]
\centering
\begin{tabular}{cc}
\epsfig{figure=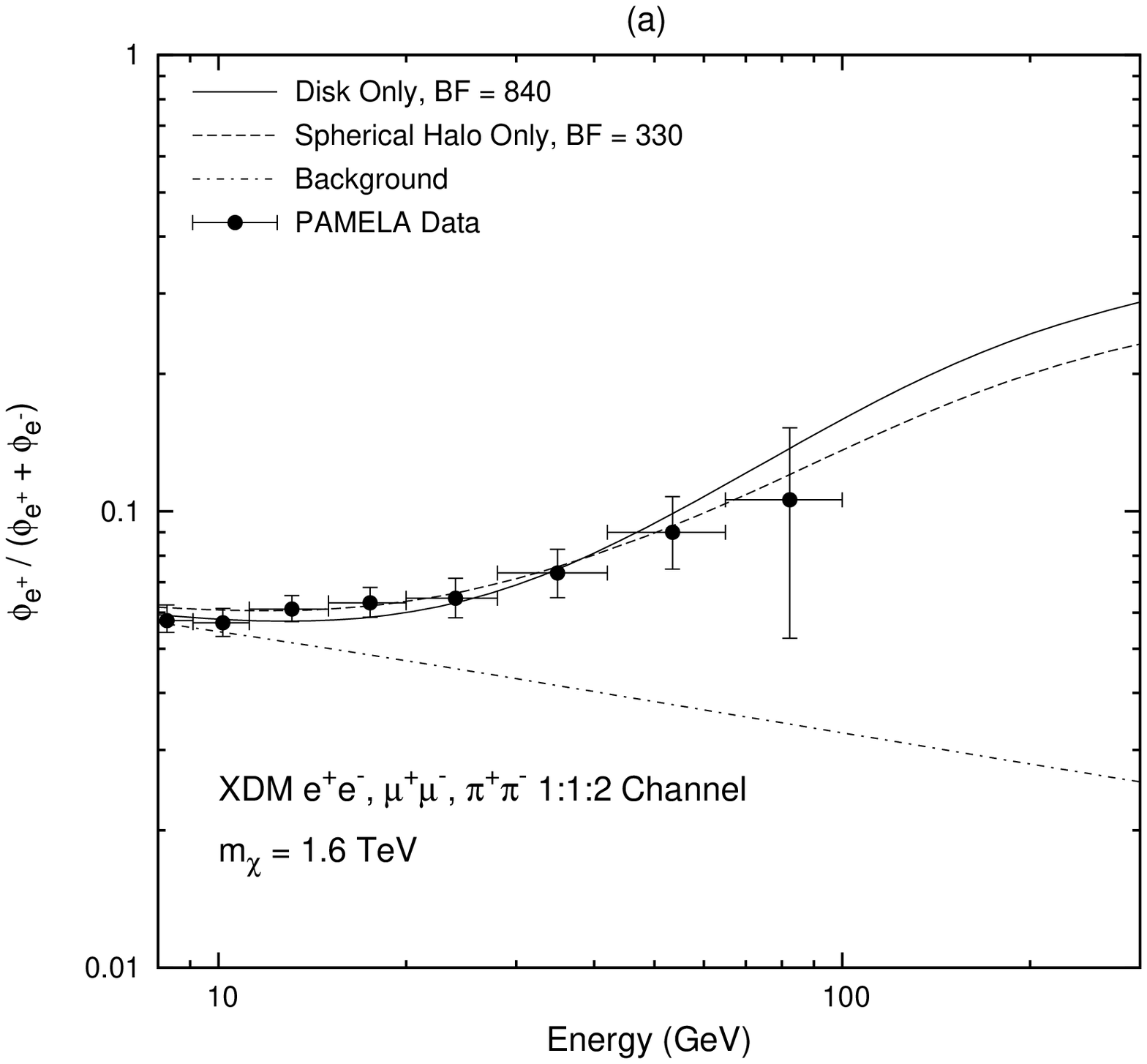,width=3.0in} &
\epsfig{figure=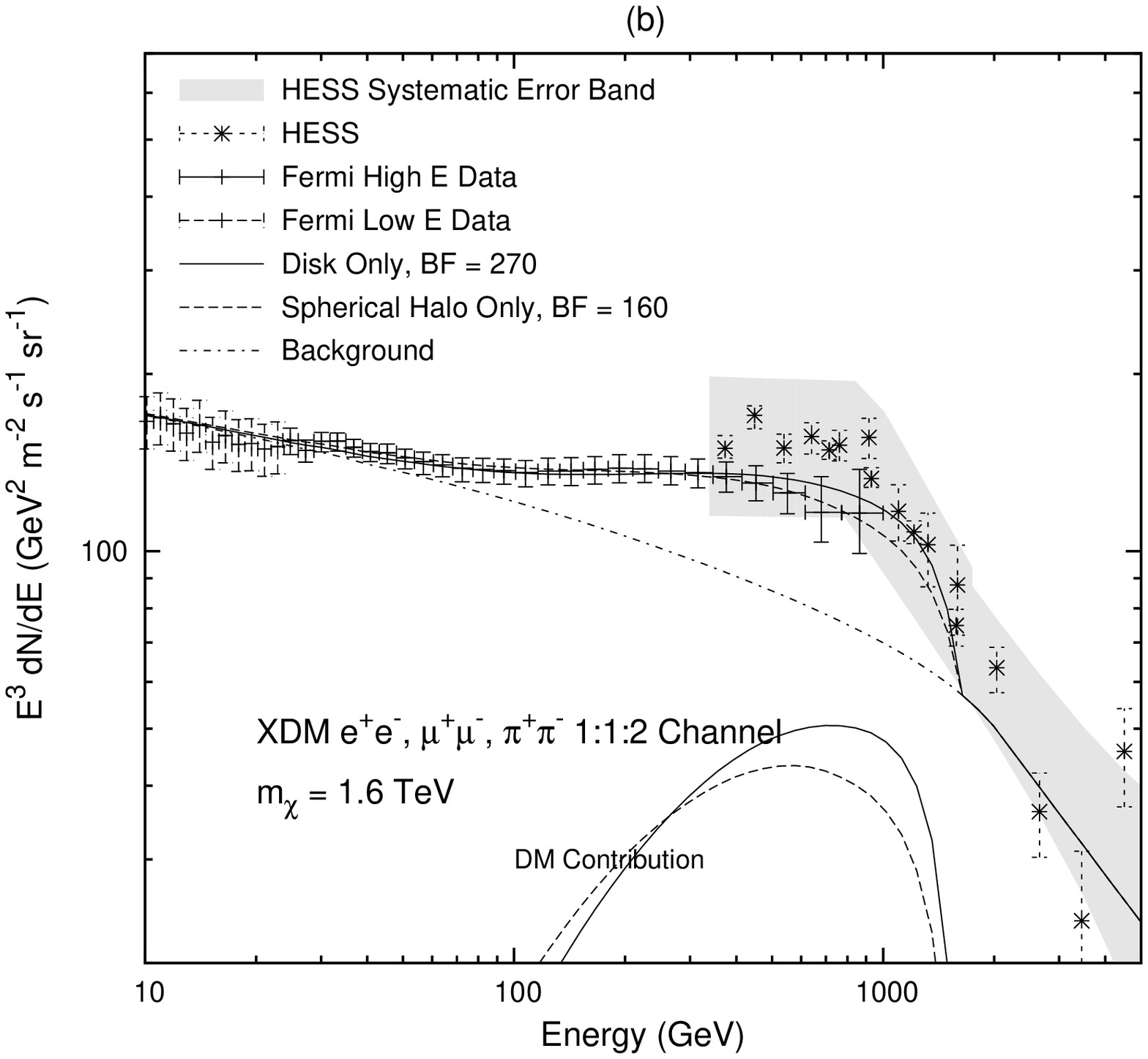,width=3.0in} \\
\end{tabular}
\caption{Cosmic ray signals as in Fig. \protect \ref{fig:XDM_singlemodes}.  XDM $e^{+}e^{-}$, $\mu^{+}\mu^{-}$ and $\pi^{+}\pi^{-}$ channels with relative BR's of 1:1:2.}
\label{fig:XDM_mixed}
\end{figure}

\begin{figure}[t!]
\centering
\begin{tabular}{cc}
\epsfig{figure=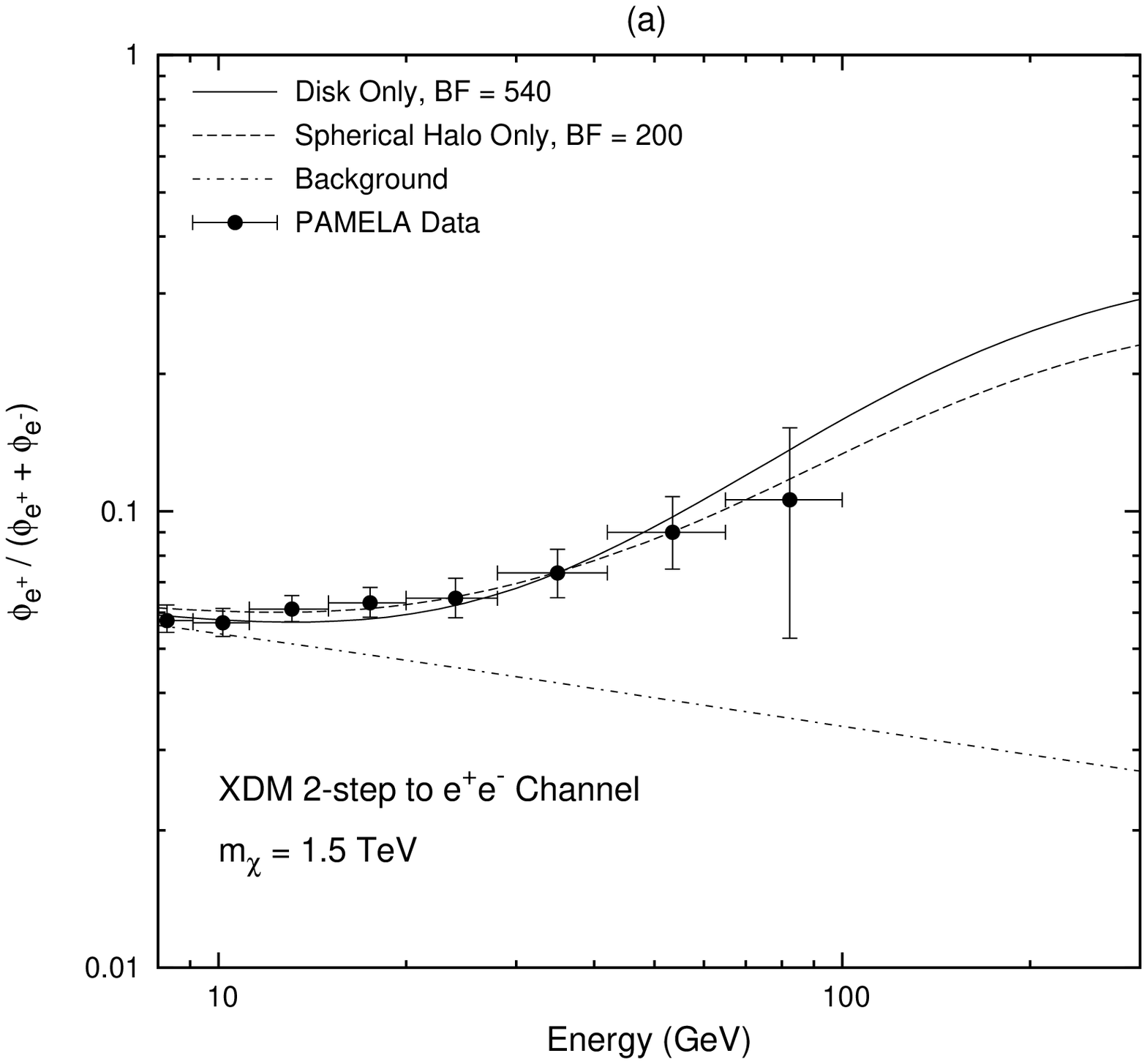,width=3.0in} &
\epsfig{figure=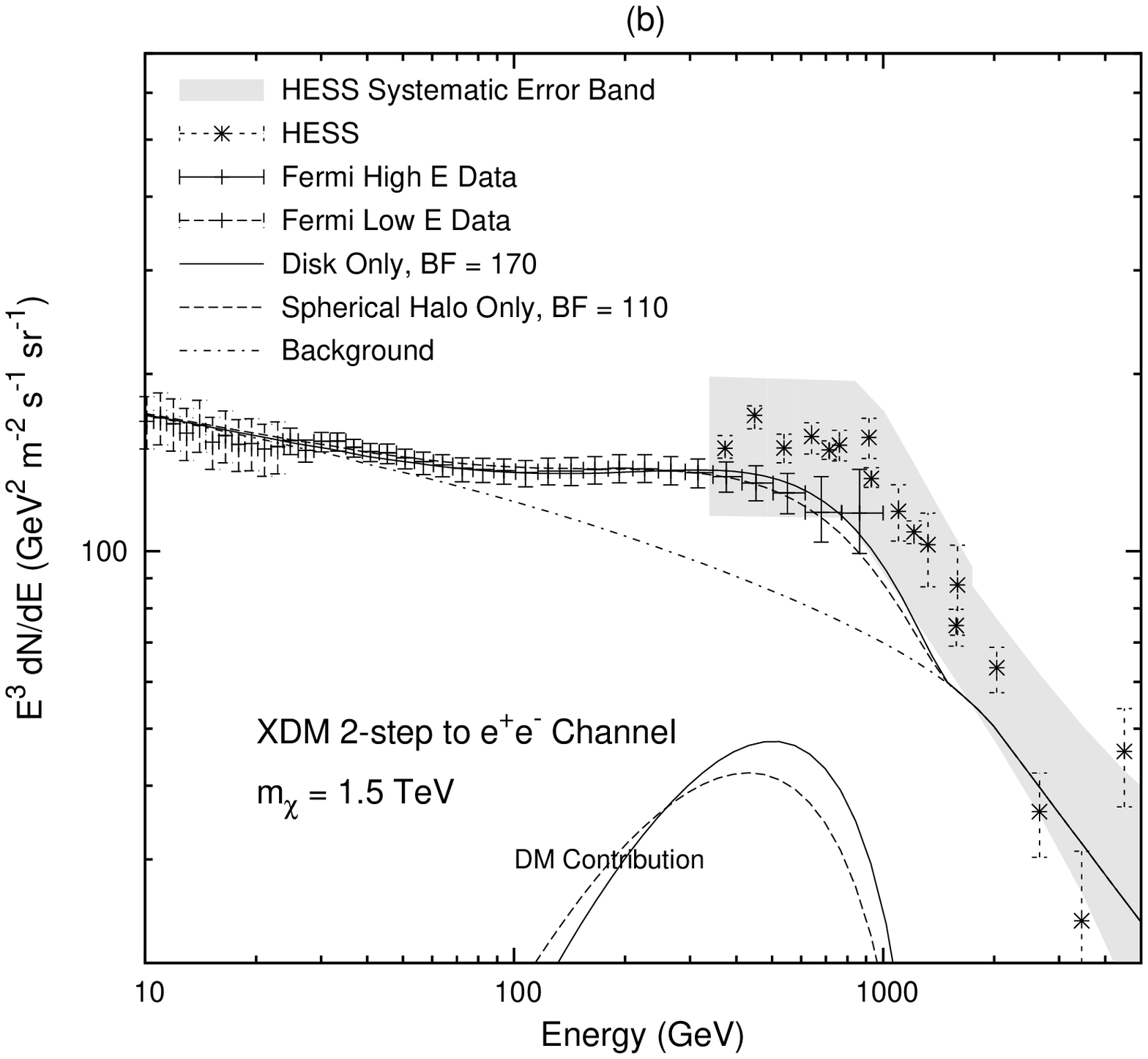,width=3.0in} \\
\vspace{5pt}
\epsfig{figure=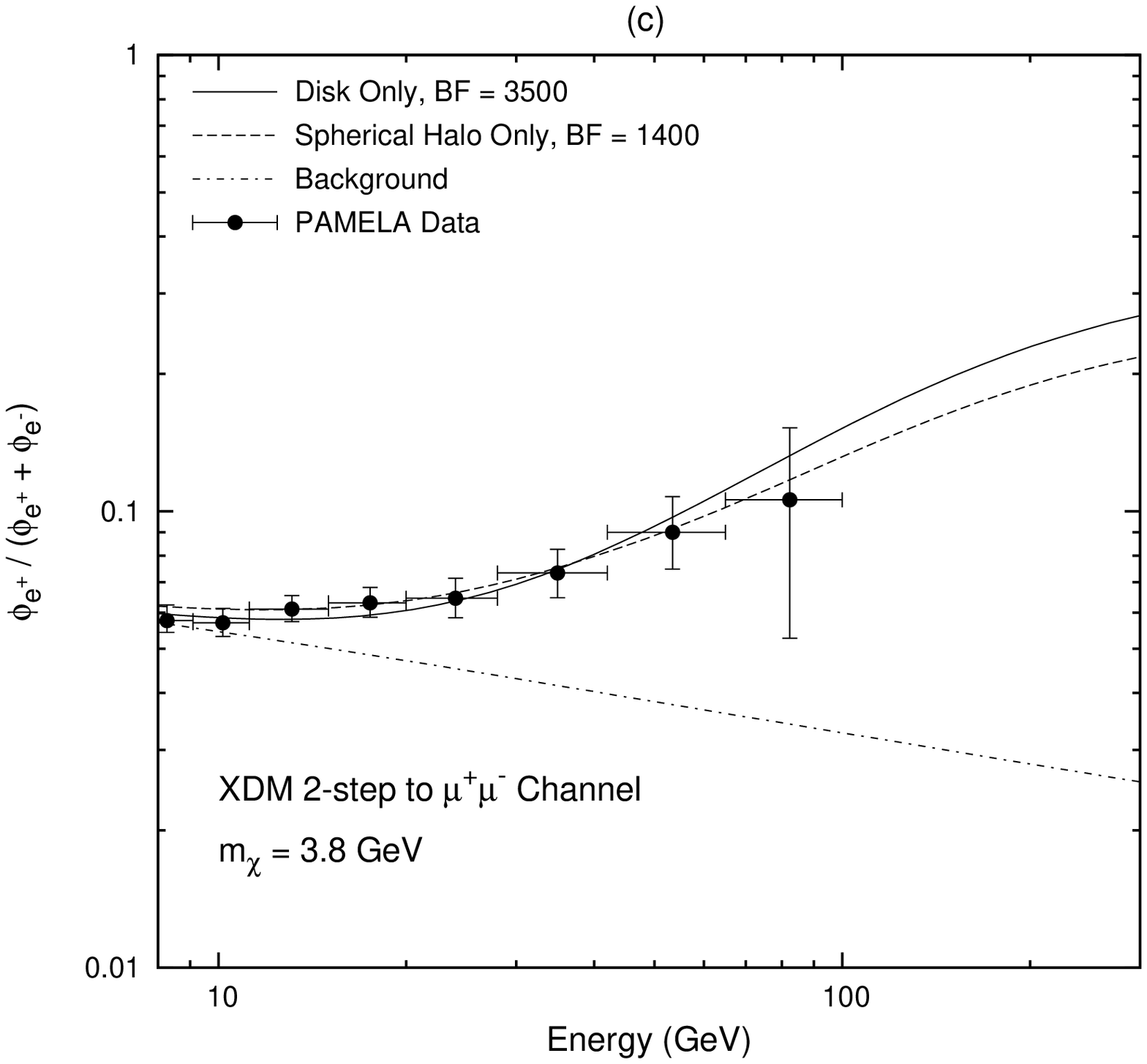,width=3.0in} &
\epsfig{figure=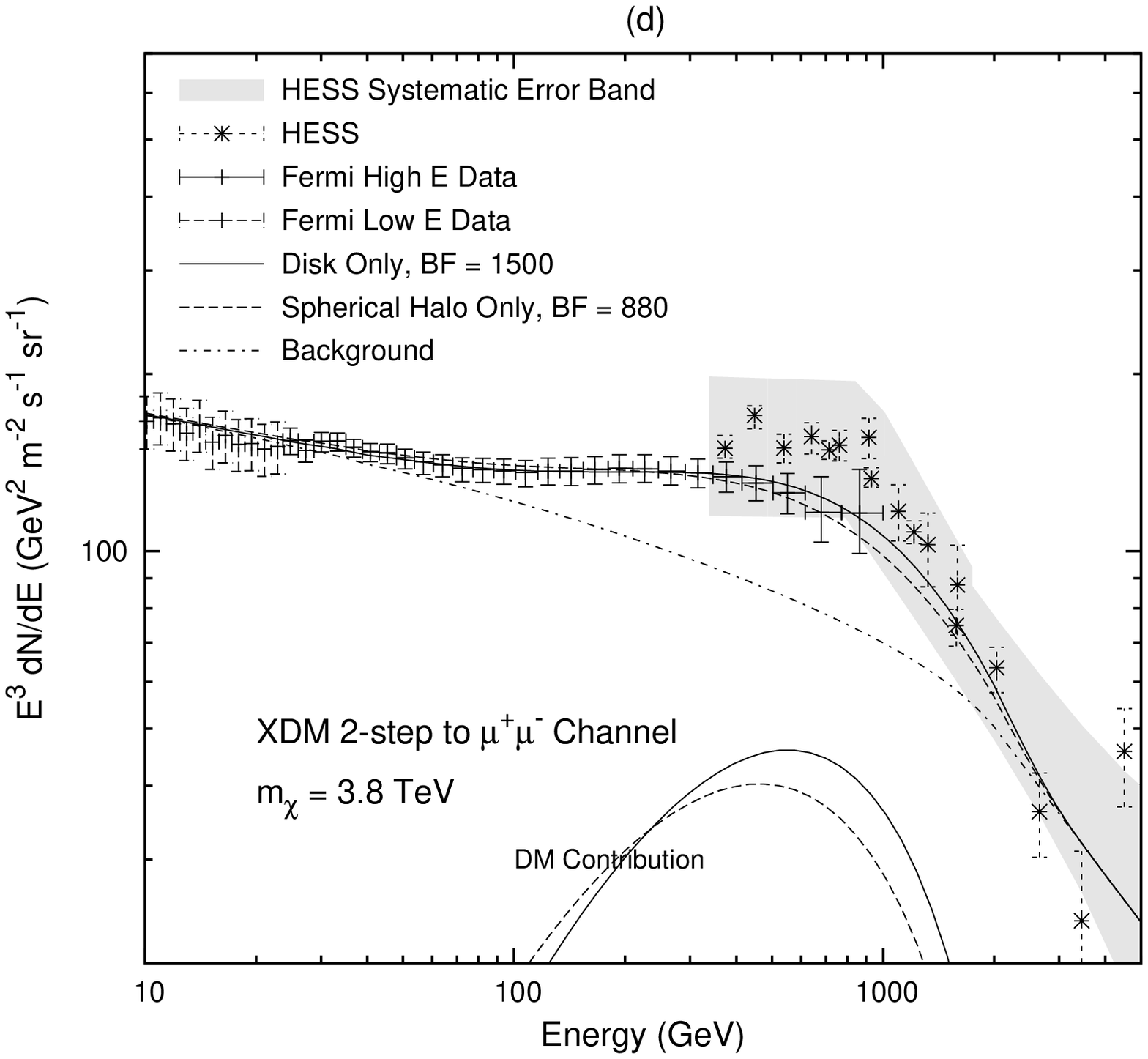,width=3.0in} 
\end{tabular}
\caption{Cosmic ray signals as in Fig.~\protect \ref{fig:XDM_singlemodes}. Top: XDM $e^{+}e^{-}$ through two steps with $m_{\Phi}\sim 10 m_{\phi}$ and $m_{\phi}<2m_{\mu}$.  Bottom: XDM $\mu^{+}\mu^{-}$ through two steps with $2m_{\mu}<m_{\phi}<2m_{\pi}$.  Two-step annihilations proceed through $\chi \chi \rightarrow \Phi \Phi$, followed by the decays $\Phi \rightarrow \phi \phi$ and $\phi \rightarrow e^{+}e^{-}$ or $\phi \rightarrow \mu^{+}\mu^{-}$.}
\label{fig:XDM_2step_muons}
\end{figure}

\begin{figure}[t!]
\centering
\begin{tabular}{cc}
\epsfig{figure=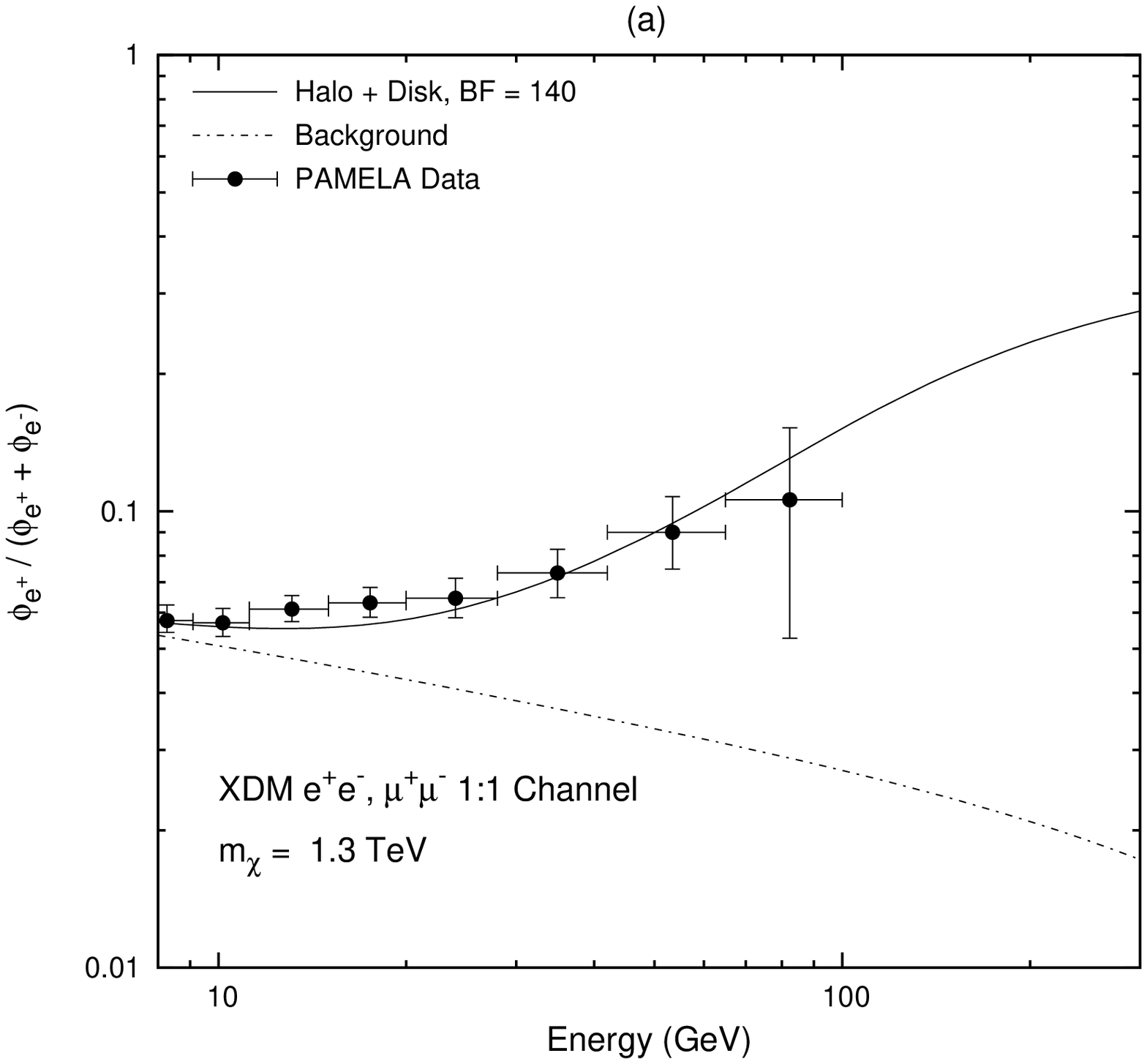,width=3.0in} &
\epsfig{figure=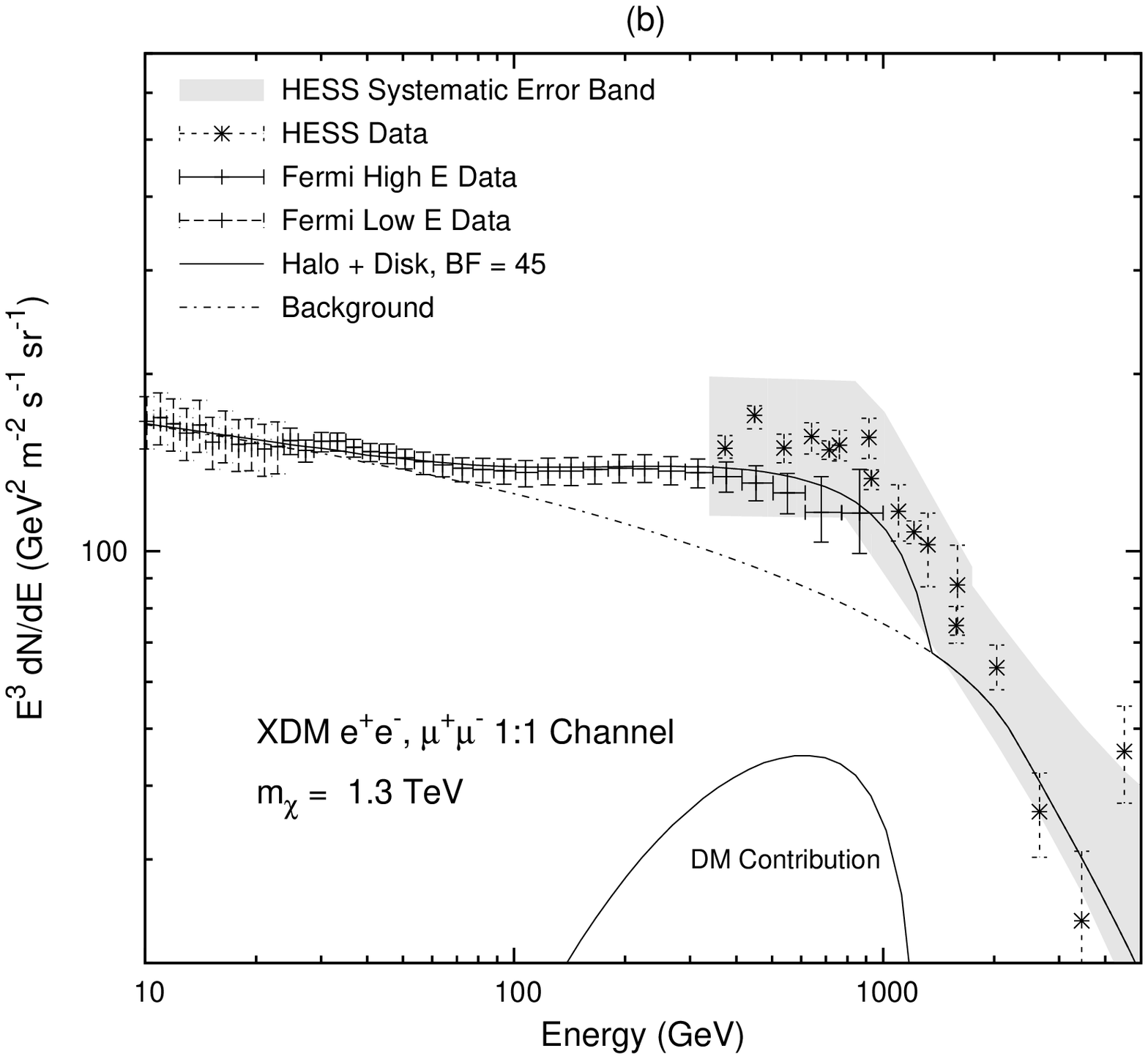,width=3.0in} \\
\end{tabular}
\caption{Cosmic ray signals as in Fig.~\protect \ref{fig:XDM_singlemodes}.  XDM $e^{+}e^{-}$ and $\mu^{+}\mu^{-}$ channels with equal BR's.  Here a combination of dark disk and Einasto profile is assumed.  The annihilation rate in the disk is taken to be 5 times the rate in the spherical halo.}
\label{fig:XDM_combo_disknhalo}
\end{figure}

\begin{figure}[t]
\centering
\begin{tabular}{cc}
\epsfig{figure=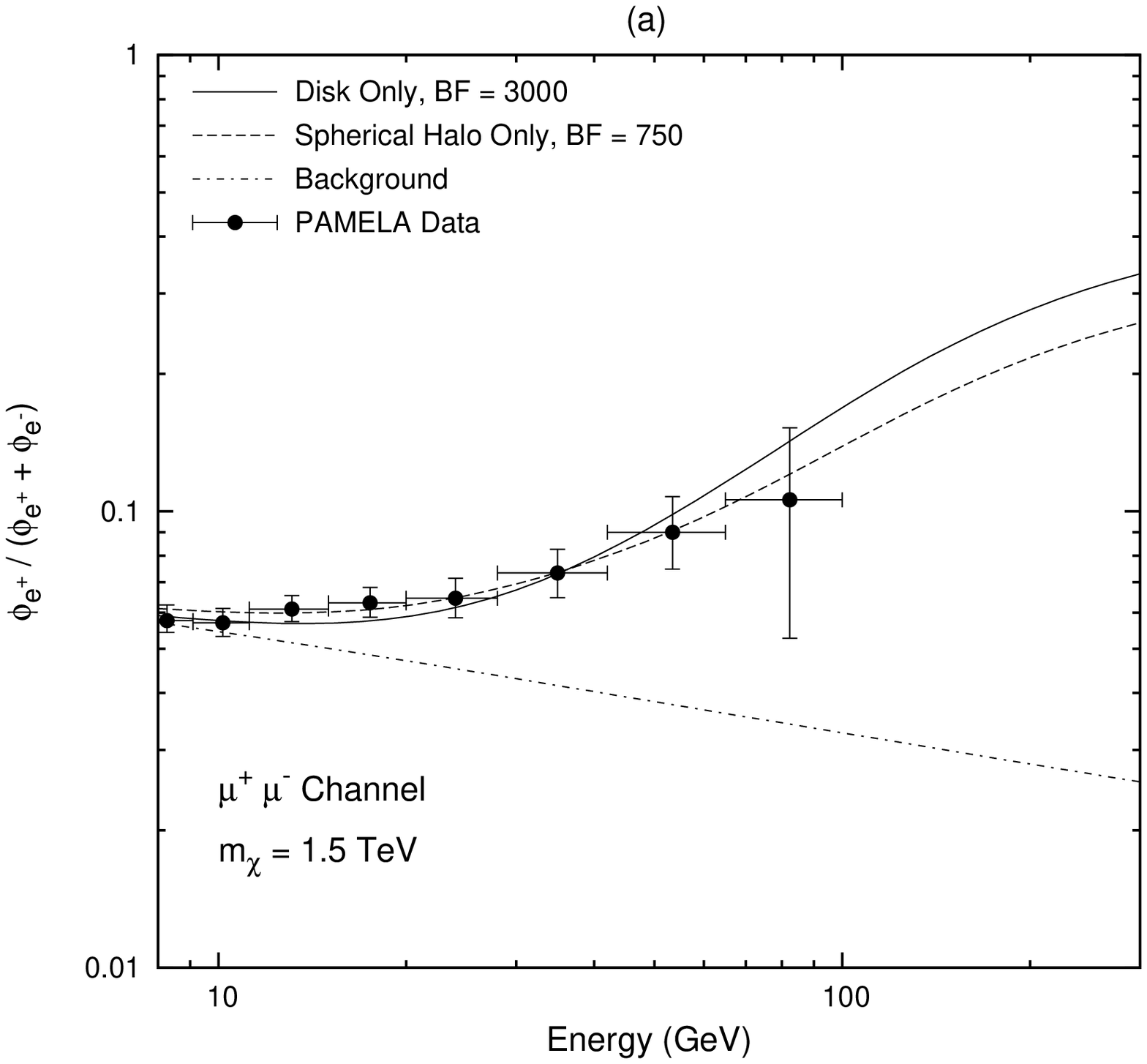,width=3.0in} &
\epsfig{figure=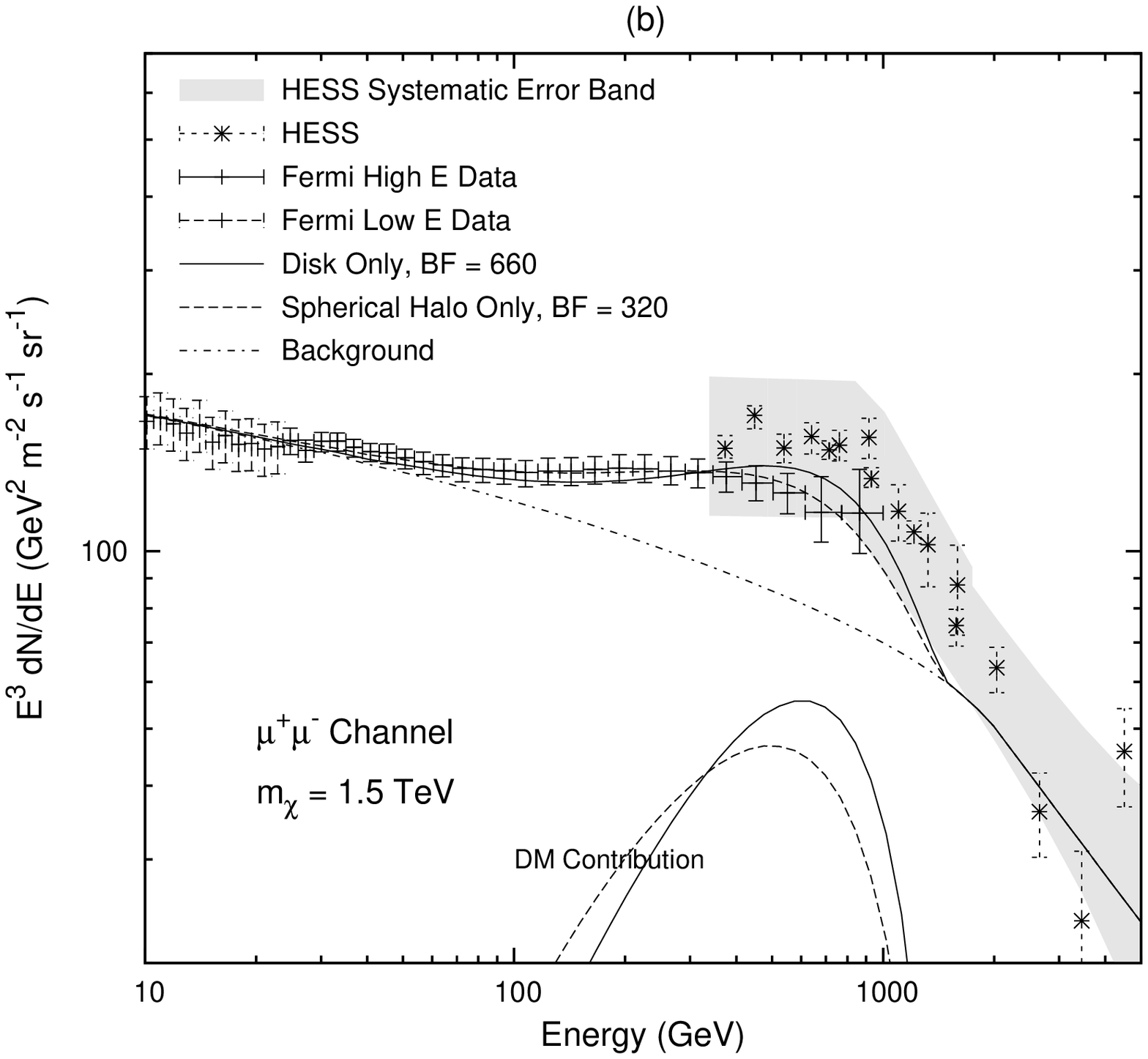,width=3.0in} 
\end{tabular}
\caption{Cosmic ray signals as in Fig.~\protect \ref{fig:XDM_singlemodes} for annihilation of DM directly to $\mu^{+}\mu^{-}$ for $m_{\chi}=1.5$ TeV.}
\label{fig:muons}
\end{figure}

In Figs.~\ref{fig:XDM_singlemodes}-\ref{fig:WW_positrons} we show the positron fraction $\Phi_{e^+}/(\Phi_{e^+}+\Phi_{e^-})$ and the $E^3$-weighted total electronic flux (the sum of the fluxes of electrons and positrons) for a number of annihilation channels.  We show the fits of the spectra to the \textit{PAMELA}, \textit{Fermi}, and HESS data separately for the dark disk and spherical halo scenarios.  The locally measured spectrum, which is composed of both a disk and a spherical halo component, is expected to lie between the two spectra shown.

For the relatively hard annihilation channels, such as $\chi\chi \rightarrow \mu^{+}\mu^{-}$ shown in Fig.~\ref{fig:muons}, the DM component of the spectrum arising from the dark disk is too peaked between 400 and 800 GeV to give really good fits to the \textit{Fermi} data, though the fits for the Einasto profile for these channels are generally good.  Those annihilation channels with much softer electron and positron spectra, for example XDM to muons or to pions only or some combination of these and XDM through a 2-step decay (see Figs.~\ref{fig:XDM_singlemodes}-\ref{fig:XDM_2step_muons}), give very good fits to the \textit{Fermi} data for both the spherical and disk profiles.  Therefore, any relative combination of disk and spherical halo contributions to the total $e^{\pm}$ flux is allowed by the \textit{Fermi} data.  (See also Section~\ref{sec:elecflux_constr} for more discussion on this).  Both DD and SH profiles give good agreement with the sharp fall-off in the spectrum seen in the HESS data, which suggests a power law of $E^{-3.9\; \pm \; 0.1}$ in the range 700 - 3500 GeV \cite{Collaboration:2008aaa}.

We note that for the flux of primary electrons we use an injection power-law of $\sim E^{-2.5}$, which agrees with conventional assumptions about the primary electron spectrum in the GeV energy range \cite{Longair, Abdo:2006fq, Kobayashi:2003kp}.  We introduce a break in the spectrum at $\sim 2$ TeV in order to have agreement with the HESS data.  Since the primary electrons composing the bulk of the background flux at TeV energies are ISM electrons accelerated from supernovae shocks, a break in the spectrum at this energy scale is natural.  However, the energy at which the break occurs depends on the most recent and closest supernovae events and is not known theoretically; rather, it comes from fitting local CR data.

The \textit{PAMELA} positron fraction data is well fit for the Einasto profile by all of the XDM annihilation channels shown. The implication for the positron fraction of the harder DM $e^{+}e^{-}$ spectra for the disk profile is a steeper rise of this fraction.  For the dark matter masses that fit the \textit{Fermi} data, $m_{\chi} \gtrsim$ 1.0 TeV, this can result in slightly worse fits to the \textit{PAMELA} data.  See, for example, Fig.~\ref{fig:XDM_singlemodes}.  However, for low DM masses that agree with \textit{PAMELA} but not \textit{Fermi}, the dark disk can improve the \textit{PAMELA} fits.

Since the velocity dispersion of the particles in the dark disk traces that of the stars in the stellar thick disk, the local values are much lower than for the spherical Einasto profile.  Typical local values of the velocity dispersion for the disk are $\sim 30 \; \rm km/s$ and for the spherical halo are $\sim 220 \; \rm km/s$.  Therefore, we expect that if the annihilation cross-section scales as $\sim 1/v$, then in the disk it can be up to $\sim 10$ higher than that of the particles in the spherical halo.  This is assuming, of course, that the Sommerfeld enhancement isn't saturated for velocity dispersions greater than that in the disk and that we are not near a resonance.  (In Section~\ref{sec:CMB_constr}, we discuss the possibility, as suggested by~\cite{Slatyer:2009yq}, that the annihilation cross-section is close to saturation for $\sigma_{v}\simeq 200\; \rm km/s$.)  In Fig.~\ref{fig:XDM_combo_disknhalo} we show an example of the \textit{PAMELA} and \textit{Fermi} spectra for a combination of disk and Einasto profiles for a disk annihilation rate that is 5 times the annihilation rate in the spherical halo.

For a specific DM mass and annihilation channel, the ratio of the boost factor needed to fit the \textit{PAMELA} data over the boost factor needed to fit the \textit{Fermi} data, $BF_{PAM}/BF_{Fermi}$, is larger for the dark disk profile than for the Einasto profile.  For example, for the XDM $\pi^{+}\pi^{-}$ channel $BF_{PAM}/BF_{Fermi} = 3.2$ for the dark disk, while for the Einasto profile the ratio is 1.9.  Due to the spectral hardening of $e^\pm$ of DD profiles compared to SH profiles, the ratio of boost factors $BF_{PAM}/BF_{Fermi}$ is expected to be higher for DD profiles than for SH profiles.  In other words, if the fluxes of $e^{+}e^{-}$ from the spherical halo and from the dark disk are equal at 100 GeV, then the flux due to the spherical component will be greater for $E\lesssim 100$ GeV, while the flux for the dark disk component will be greater for $E\gtrsim 100$ GeV.  Thus, a spectral hardening would allow annihilation channels that under-produce at the highest \textit{Fermi} energies in a spherical halo to have improved fits to the data for a disk halo.  On the other hand, the ``flatness" of the \textit{Fermi} data (in $E^3$-weighted plots) can be used to constrain the disk contribution to the overall flux.  Additionally, \textit{WMAP} data can put upper limits on the saturated Sommerfeld enhancement \cite{Slatyer:2009yq}, so the contribution of the dark disk for various channels can be constrained in this way as well.  See Section~\ref{subsec:boostfactorconstraints}.

\vspace{\baselineskip}
\noindent{\textbf{Non-Sommerfeld Enhanced Channels}}

The annihilation channels that proceed through one or more light mediators (see Figs.~\ref{fig:XDM_singlemodes}-\ref{fig:XDM_combo_disknhalo}) require large boost factors to fit the \textit{PAMELA} and \textit{Fermi} data, and these can be explained by the Sommerfeld enhancement.  Annihilation channels directly to leptons also require large boost factors, $BF\sim 500$ for the spherical halo, but in these scenarios there is no long range force to give rise to the Sommerfeld enhancement needed to motivate such a large boost factor.  A Breit-Wigner resonance, as suggested by \cite{Guo:2009aj}, could account for the large annihilation rate needed at $z=0$ to produce the signals we observe today, while not impacting the thermal relic density at freeze-out.  Also, WIMPs produced non-thermally, such as those of \cite{Dutta:2009uf, Bi:2009am}, can naturally have large annihilation cross-sections that give good fits to the \textit{PAMELA} and \textit{Fermi} data.  For this reason, we consider here such non-Sommerfeld enhanced scenarios in the context of a dark disk.

In Fig.~\ref{fig:muons} we show the positron fraction and total electronic flux for dark matter annihilating monochromatically to muons, $\chi \chi \rightarrow \mu^{+} \mu^{-}$.  Good fits to both the \textit{PAMELA} and \textit{Fermi} data can be achieved with the spherical halo profile.  As mentioned earlier, the disk profile gives an electronic spectrum that is too hard to fit well the \textit{Fermi} data, though it does a good job reproducing the positron fraction.  Additionally, the ratio $BF_{PAM}/BF_{Fermi}$ is 4.5 for the dark disk, which suggests that a dark disk dominant scenario does not accurately explain both signals simultaneously.

While direct decay to taus, $\chi \chi \rightarrow \tau^{+} \tau^{-}$, gives good fits to the \textit{Fermi} and \textit{PAMELA} data, the annihilation rate for the tau channel is highly constrained by the \textit{Fermi} diffuse $\gamma$-ray data \cite{Cirelli:2009dv, Papucci:2009gd} and by the HESS measurements of gamma rays in the Galactic Ridge \cite{Aharonian:2006au, Meade:2009rb}.   (The $\tau^{+}\tau^{-}$ channel produces a high energy prompt component of gamma rays through the $\pi^{0}$s created in the decay chain.)  Annihilation directly into $e^{+}e^{-}$ gives a propagated spectrum $e^{+}e^{-}$ with spectral index of -2 and a hard cut-off at $E=m_{\chi}$ \cite{Cholis:2008wq, Malyshev:2009tw}; such a spectrum does not fit the \textit{Fermi} and HESS $e^{-} + e^{+}$ data well.

\subsection{Antiprotons}\label{subsec:antiprotons}

\begin{figure}[t!]
\centering
\begin{tabular}{cc}
\epsfig{figure=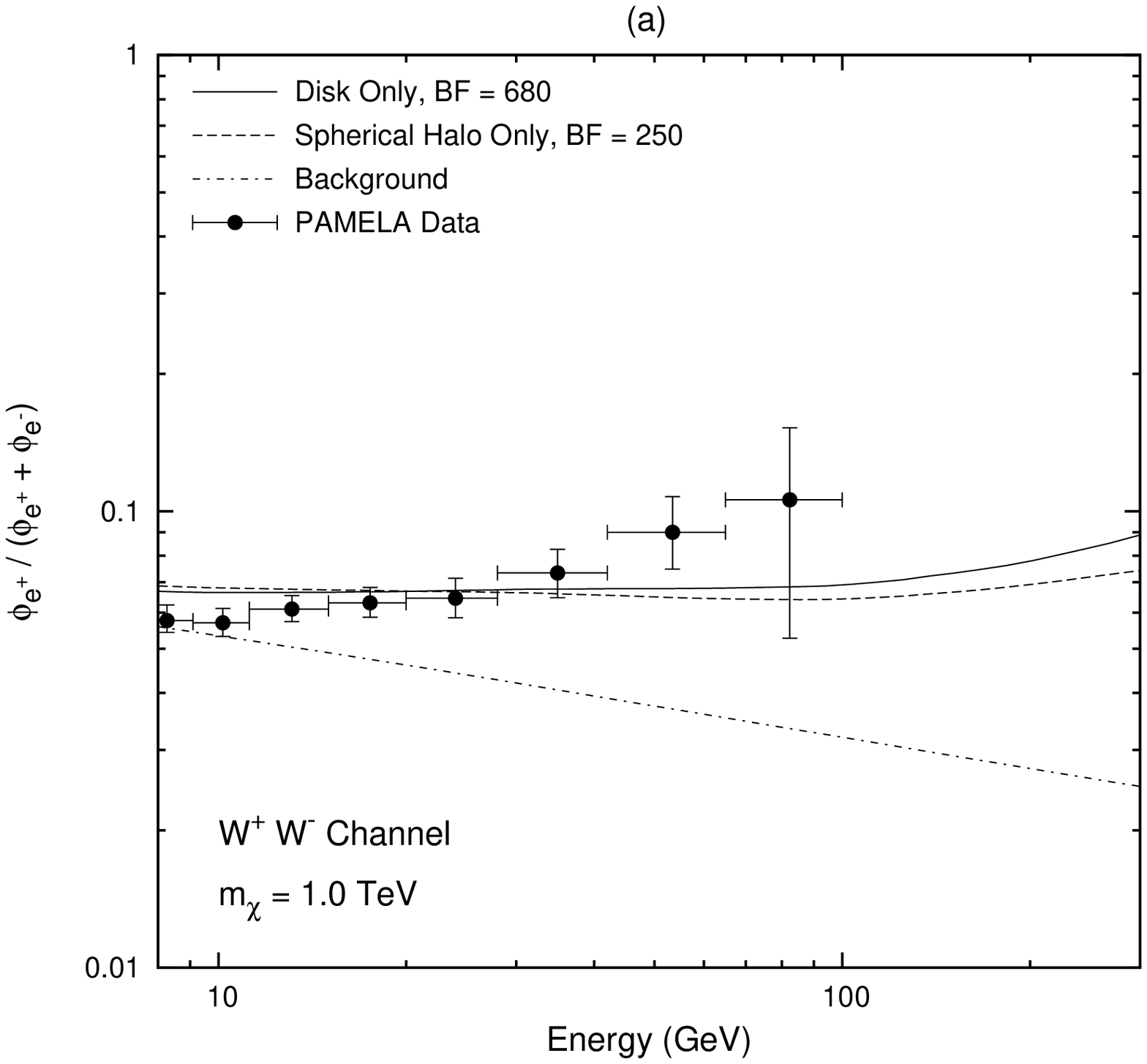,width=3.0in} &
\epsfig{figure=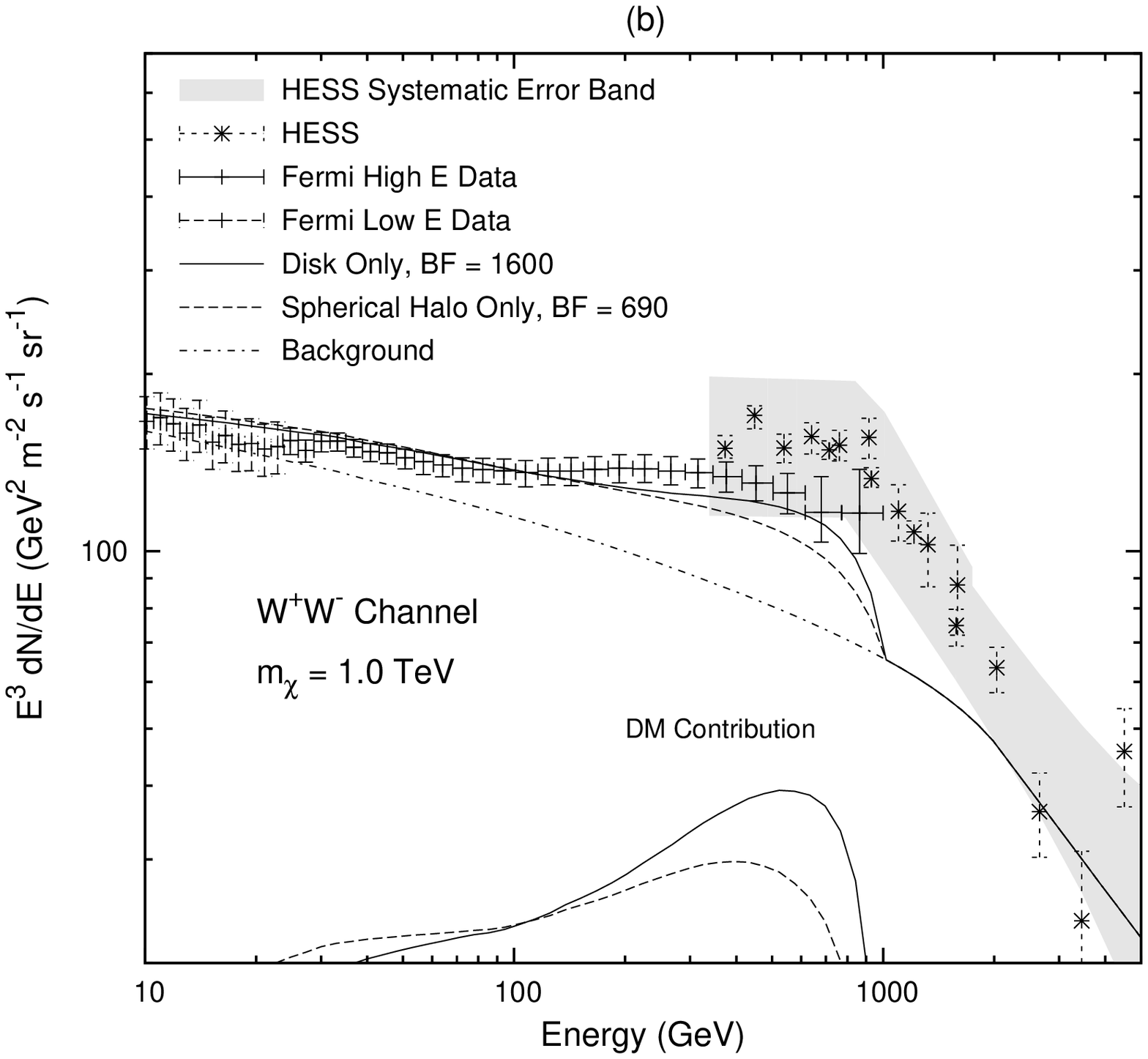,width=3.0in} 
\end{tabular}
\caption{Cosmic ray signals as in Fig.~\protect \ref{fig:XDM_singlemodes} for DM annihilation to $W^{+}W^{-}$ with $m_{\chi}=1.0$ TeV.}
\label{fig:WW_positrons}
\end{figure}

\begin{figure}[t!]
\centering
\begin{tabular}{cc}
\epsfig{figure=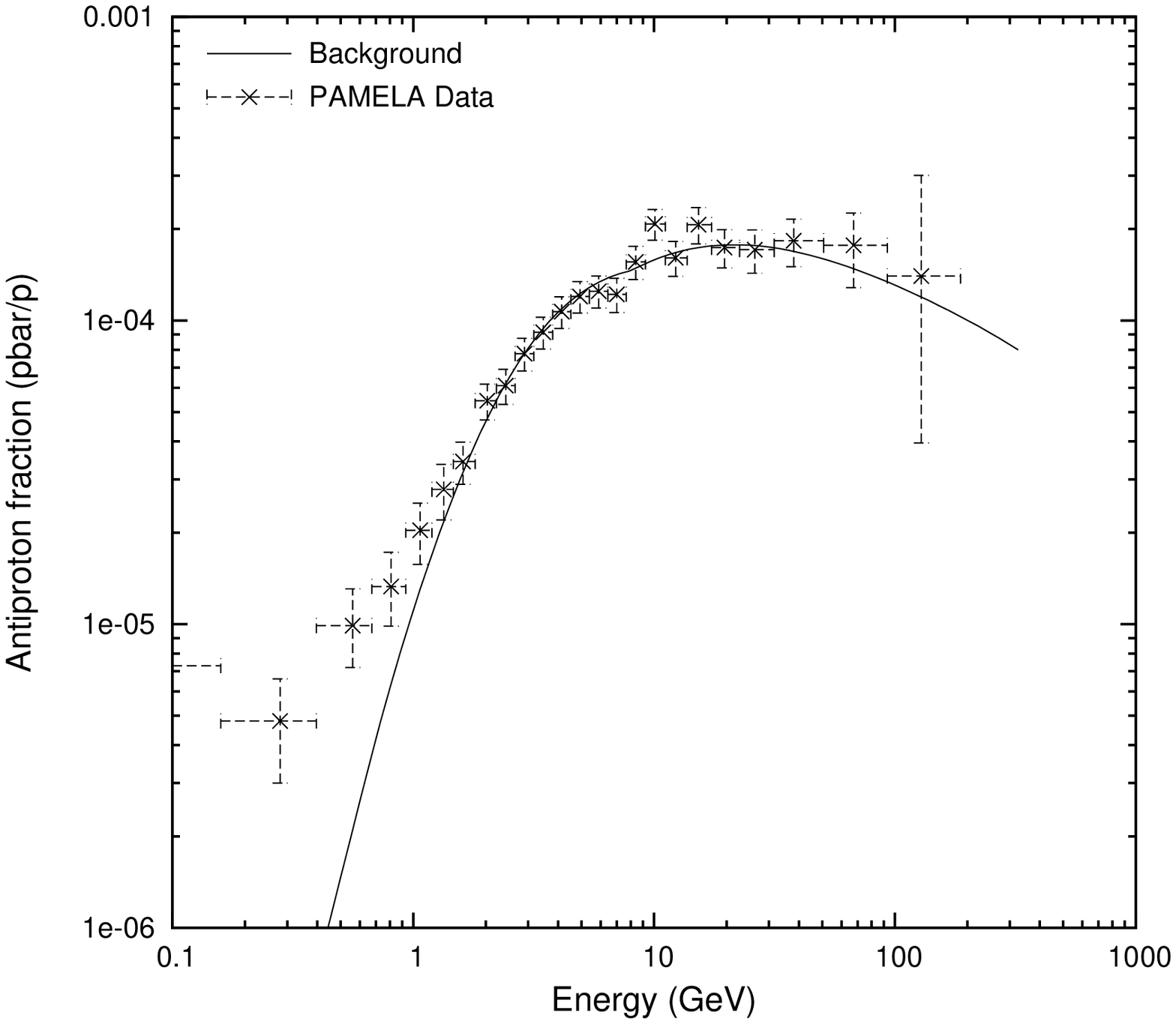,width=3.0in} &
\epsfig{figure=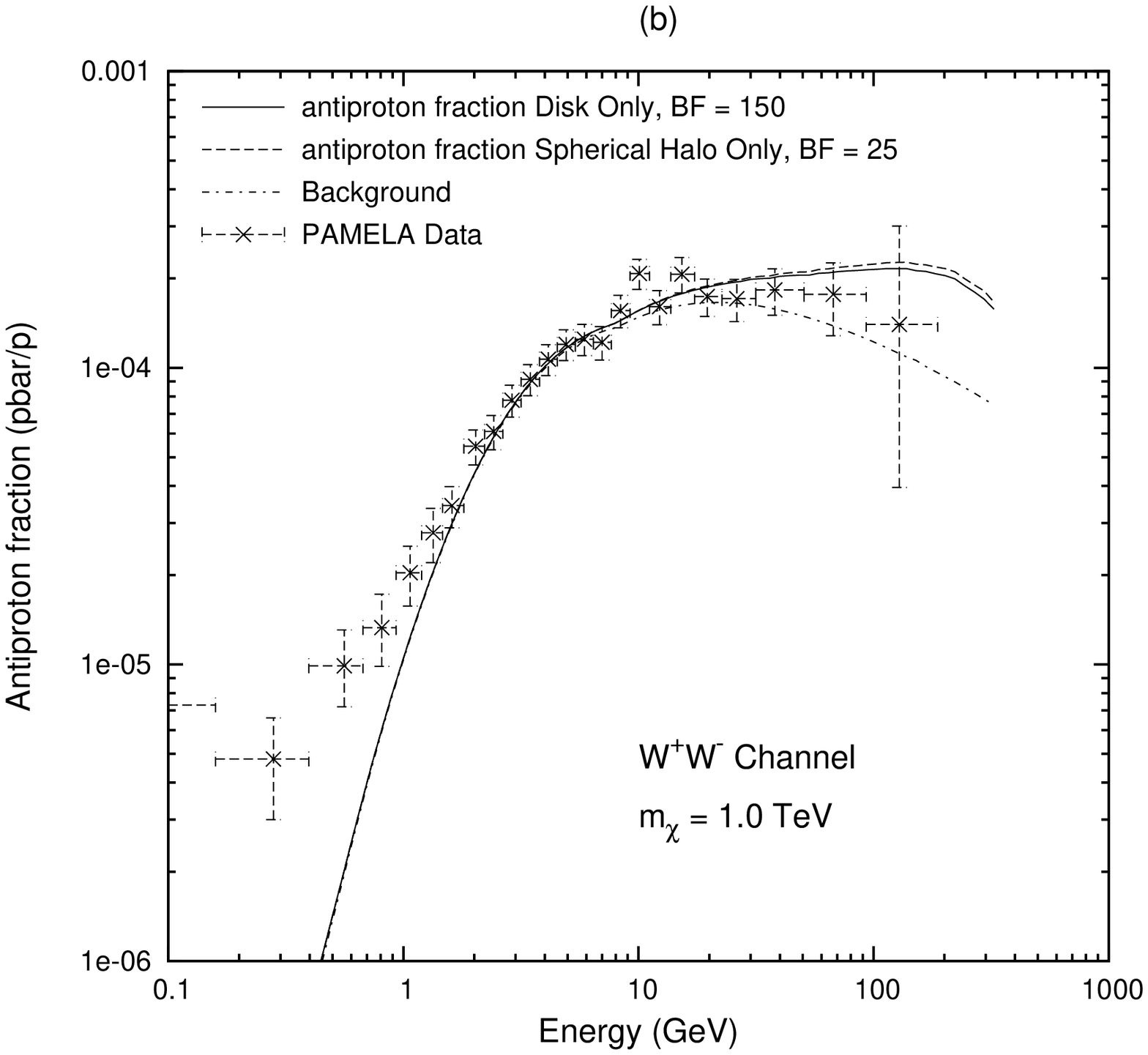,width=3.0in} 
\end{tabular}
\caption{Antiproton over proton ratio $\bar{p}/p$ as a function of energy.  Left: Background for the propagation and diffusion parameters used in our calculations. Right: For DM annihilating through $W^{+}W^{-}$ with $m_{\chi}=1.0$ TeV for the Dark Disk profile (\textit{solid}) and for the Einasto profile (\textit{dashed}).  Here the background (\textit{dot-dashed}) is smaller than the standard background by $10\%$.  Data of the antiproton to proton flux ratio are from \cite{:2010rc}.}
\label{fig:WW_antiprotons}
\end{figure}

Annihilation through $W^{+}W^{-}$ gives very poor fits to the \textit{Fermi} and \textit{PAMELA} data.  Most of the electrons and positrons are produced in the decay chains of the hadrons and taus, so they have relatively soft spectra.  The resulting abundance of low energy $e^{+} e^{-}$ pairs gives a positron fraction that is much too flat and an $e^{+} + e^{-}$ spectrum that doesn't simultaneously agree with the \textit{Fermi} data around 20 GeV and the data around 1 TeV.  In Fig.~\ref{fig:WW_positrons} we show the results for $m_{\chi} = 1$ TeV, the DM mass that best fits the \textit{Fermi} data.  The fit to the \textit{Fermi} data is significantly better for the dark disk profile than for the spherical halo. Additionally, the fit of the positron fraction to the \textit{PAMELA} data has a modest improvement.  Thus, for the $W^{+}W^{-}$ channel the favored scenario is one in which the dark disk is the dominant contribution to the local DM density.

An additional benefit of the dark disk for the $W^\pm$ channel is that the constraint placed on the annihilation rate by the \textit{PAMELA} antiproton data is somewhat relaxed.  In Fig.~\ref{fig:WW_antiprotons}b we show that a very similar antiproton ratio is obtained for the dark disk with $BF=150$ as for the Einasto profile with $BF=25$. The factor of $\sim 6$ ``improvement" is a reflection of the fact that the $p$'s propagate to us with little energy loss from the entire diffusion zone, and the disk has a smaller average dark matter density throughout the diffusion zone, as discussed in Section~\ref{sec:intro}.  This, of course, depends greatly on the relative widths of the diffusion zone and the dark disk.  For diffusion zones with widths much larger than the scale width of the disk, the improvement becomes significant, i.e. factors are much larger than 1.  For narrow diffusion zones such as the $\sim1$ kpc suggested by \cite{Hooper:2009fj}, the thickness of the disk is irrelevant, as it is thicker than the diffusion zone.  If the dark disk is thicker than the diffusion zone, improvement can still be had if the average value of the DM density over the diffusion zone is less for the dark disk than for a spherical halo profile.  We note that for $m_{\chi}=1.0$ TeV annihilating through $W^{+}W^{-}$ in the dark disk only scenario, the BF needed to fit the \textit{PAMELA} positron fraction is still $\sim 4$ times larger than that allowed by the \textit{PAMELA} $\bar{p}/p$ ratio.  For comparison, in the spherical halo only scenario the BF for the \textit{PAMELA} positron fraction is $\sim 10$ times larger than that allowed by $\bar{p}/p$.  However, since the fits to the \textit{PAMELA} data are so poor, the $BF$'s should not be taken too seriously.  In summary, the discrepancy between the larger boost factors needed to fit the positron fraction data and the smaller boost factors needed to fit the antiproton fraction data can be reduced by including a disk component.  Thus, a higher branching ratio to hadronic channels is allowed for a dominantly dark disk profile.

\subsection{Constraints on Boost Factors}\label{subsec:boostfactorconstraints}

\vspace{\baselineskip}
\noindent{\textit{Constraints from the Fermi $e^{+}+e^{-}$ flux}}\label{sec:elecflux_constr}

The combined flux from a spherical halo and a dark disk can be written as\footnote{The $\frac{dN_{comb}}{dE}$ comes from the term $\propto \rho_{DD}*\rho_{SH}$ in Eq.~\ref{eq:annih_rate_Somm}, where we assume that the relative velocity dispersion between the particles in the DD and the SH is $\sim 200$ km/s.}

\begin{equation}
\label{eq:total_eflux}
\frac{dN_{total}}{dE} = \frac{dN_{prim}}{dE}+\frac{dN_{sec}}{dE} + \frac{BF_{SH}}{(1+M)^2}\left(\frac{dN_{SH}}{dE} + R M^2 \frac{dN_{DD}}{dE} + 2 M \frac{dN_{comb}}{dE}\right).
\end{equation}

\noindent Here $M$ is the ratio of the local DM densities, $\rho_{0_{DD}}/\rho_{0_{SH}}$, $R$ is the ratio of the thermally-averaged annihilation cross-sections, $R = \langle\sigma\vert v\vert\rangle_{DD}/\langle\sigma\vert v\vert\rangle_{SH}$, and $BF_{SH}$ is the Sommerfeld enhancement in the spherical halo (i.e. for the velocity dispersion in the spherical halo).

Using the $e^{+}+e^{-}$ flux as measured by \textit{Fermi}, we can constrain $M$, $R$, and $BF_{SH}$.  Since the \textit{Fermi} data constrain the total flux of Eq.~\ref{eq:total_eflux}, constraints on any of these parameters can be used as further input to constrain the rest of the parameters.  We consider the following values for the parameters:
\begin{equation}
\label{eq:init_contraints}
	 0 < M < 1 \textrm{,} \hspace{.25in} \rho_{0_{DD}} \leq \rho_{0_{SH}} \textrm{,} \hspace{.25in} \rho_{0_{SH}} + \rho_{0_{DD}} = 0.4\;\rm GeV cm^{-3}.
\end{equation}

We take $R_{max}=100$ to be the maximum ratio of the annihilation cross-section of DM particles in the disk halo to the spherical halo.  This value could occur if we live close to a resonance, where the enhancement scales as $1/v^{2}$, since the velocity dispersion in the spherical halo is $\sim 220\; \rm km/s$ and in the dark disk halo is $\sim 30\; \rm km/s$.\footnote{Since we want to place an upper bound on the Sommerfeld enhancement in the dark disk, we use a value for the velocity dispersion that is on the low side, but within the range suggested by simulations \cite{Read:2008fh}.  Such a value is motivated in dominant thin disk scenarios.}  This assumes that the Sommerfeld enhancement does not saturate at velocities above $\sim 30\; \rm km/s$.  If saturation occurs for annihilation in the dark disk but not in the spherical halo, i.e. at a velocity between $\sim 220$ and $\sim 30$ km/s, then $R$ is the ratio of the saturated cross-section in the disk to the cross-section in the spherical halo.

In our calculations we allow the normalization of the primary $e^{-}$ flux to vary in the range $1.48-2.04 \times 10^{-2} \; \rm GeV^{-1} m^{-2} s^{-1} sr^{-1}$ at 20 GeV, and the power law index ($dN/dE \sim E^{-\alpha}$) to vary in the range $3.2<\alpha<3.35$.  Since the secondary $e^{\pm}$'s contribute negligibly to the total $e^{+}+e^{-}$ flux for $E>20$ GeV, we hold them constant assuming $dN_{sec}/dE\sim E^{-3.7}$ and a normalization of $6.7 \times 10^{-4} \; \rm GeV^{-1} m^{-2} s^{-1} sr^{-1}$ at 20 GeV.

In column 4 of Table~\ref{tab:Table 95CL} we show the 95$\%$ C.L. values for $R$, the ratio of the annihilation cross-section of DM particles in the disk to the spherical halo, assuming that the local DM density receives equal contributions from the disk and the spherical halo.  If the \textit{Fermi} errorbars are correlated, the $\chi^{2}$ distribution test is not an exact measure to calculate confidence levels.  Since we don't have any information apart from the errorbars, the mean values of the fluxes, and the energy binning, we consider each point as an independent normally-distributed variable with mean and standard deviation given by the \textit{Fermi} data.  For all cases except the XDM electron channel, the channel with the hardest DM $e^\pm$ spectra, the annihilation rate in the disk can be at least 100 times the annihilation rate in the spherical halo, indicating that the \textit{Fermi} data is consistent with the presence of a dark disk component in a non-saturated resonance scenario.  In column 5 of Table~\ref{tab:Table 95CL} we show the 95$\%$ C.L. for the maximum value of $M$, the ratio of the local DM densities, assuming that the annihilation rate in the disk is 100 times the annihilation rate in the spherical halo.  The \textit{Fermi} $e^{+}+e^{-}$ data does not constrain the ratio of the local DM densities within our allowed range of values (0 to 1), except for the XDM electron channel.  The results of columns 4 and 5 in Table~\ref{tab:Table 95CL} suggest that for the XDM channels giving the best fits to \textit{Fermi} and \textit{PAMELA}, the dark disk may be a dominant contributor to the local flux of $e^\pm$ coming from dark matter annihilation.  In the latter case, the Sommerfeld enhancement in the spherical halo is typically smaller,by an order of magnitude, than the enhancement when the spherical halo is the dominant contribution.

\begin{table}[t]
\begin{center}
\begin{tabular} {|l||c|c|c|c|c|}
\hline
 Channel & $M_{\chi}$(TeV) & $BF_{SH}$ & $R_{max} (M=1)$ & $M_{max} (R=100)$ & $BF_{WMAP5}$ \\
\hline\hline
XDM to $e^{\pm}$ & 1.2 & $10-160$ & 90 & 0.9 & 210 \\
\hline
XDM to $\mu^{\pm}$ & 2.5 & $30-1100$ & 100 & 1 & 1300 \\
\hline
XDM to $\pi^{\pm}$ & 3.1 & $40-2000$ & 100 & 1 & 1900 \\
\hline
XDM 1:1:2 & 1.6 & $7.0-380$ & 100 & 1 & 560 \\
\hline
XDM 2-step to $e^{\pm}$ & 1.5 & $2.0-250$ & 100 & 1 & 260 \\
\hline
XDM 2-step to $\mu^{\pm}$ & 3.8 & $10-2100$ & 100 & 1 & 1900 \\
\hline
$\mu^{\pm}$ & 1.5 & $220-1550$ & 1* & 1 (R = 1) &750 \\
\hline
\end{tabular}
\end{center}
\caption{Table of allowed values of the boost factor for the Einasto profile for fits to the \textit{Fermi} data, $BF_{SH}$, the maximum allowed values of $R = \langle\sigma\vert v\vert\rangle_{DD}/\langle\sigma\vert v\vert\rangle_{SH}$ assuming $M=\rho_{0_{DD}}/\rho_{0_{SH}}=1$ (95$\%$ C.L.), the maximum allowed values of $M$ for $R=100$, and the boost factors $BF_{WMAP5}$ excluded by $\textit{WMAP}5$ data at 95$\%$C.L.  *There is no Sommerfeld enhancement for $\chi \chi \rightarrow \mu^+ \mu^-$.}
\label{tab:Table 95CL}
\end{table}

In the limiting case, $\rho_{0_{DD}}/\rho_{0_{SH}}\ll 1$, the dark disk contribution can be significant to the flux of $e^\pm$ locally only if the masses of the dark matter $m_{\chi}$ and the mediator $m_{\phi}$ and the value of the coupling between the two $\lambda$ are such that for a velocity dispersion of $\sim 30 \;\rm km/s$ the annihilation cross-section has a resonance \cite{ArkaniHamed:2008qn, Pieri:2009zi}.  The results shown in Table~\ref{tab:Table 95CL} indicate that XDM channels that produce $e^{\pm}$ with a spectral index $\alpha > 2$ at 300 GeV\footnote{XDM to pions only, XDM 2-step to muons, and XDM to $e^{\pm}$, $\mu^{\pm}$ and $\pi^{\pm}$ with relative ratios of 1:1:2}, where $\frac{dN}{dE}\propto E^{-\alpha}$, are allowed within 95$\%$ C.L. to have a resonance in their cross-section.  We clarify that in the cases of resonant Sommerfeld enhancement at $z=0$ for a velocity dispersion of $30$ km/s, it is the dark disk that gives the dominant contribution to the local $e^{\pm}$ flux, while the spherical halo is constrained to be subdominant.  See Table~\ref{tab:Table 95CL}.  At $v_{rel} \sim 200$ km/s for $m_{\chi}\sim$ TeV, the Sommerfeld enhancement is $\mathcal{O}(10)$.

Since the \textit{PAMELA} positron fraction has only 6 data points at energies $E_{e^\pm} > 10$ GeV, it constrains the annihilation rates less than the \textit{Fermi} $e^{+}+e^{-}$ flux. Additionally, the fact that the secondary $e^{+}$'s are still at least half of the total $e^{+}$ flux for  $E_{e^{+}} < 50$ GeV allows for greater uncertainty in the necessary DM annihilation $BF$'s.

\vspace{\baselineskip}
\noindent{\textit{Constraints from WMAP5 data}}\label{sec:CMB_constr}

The high energy electrons and positrons produced by dark matter annihilation in the early universe heat and ionize the photon-baryon plasma and thus can give rise to an increase in the ionization fraction of the plasma after recombination.  Perturbations in the ionization history around the time of last scattering ($z\sim 10^{3}$) can affect the temperature and polarization power spectra of the CMB.  \cite{Slatyer:2009yq} (see also \cite{Zavala:2009mi}) calculated the effect of annihilating DM on the plasma and placed constraints on the maximum annihilation cross-section using the \textit{WMAP}5 data.

The upper bounds on the cross-section depend linearly on the mass of the DM candidate and are inversely proportional to the efficiency $f$ with which the energy from DM annihilation gets converted into the energy of the $e^{\pm}$ products.\footnote{The efficiency $f$ depends on the annihilation channel, the redshift, and also has a weak dependence on the mass of the DM candidate.} In column 6 of Table~\ref{tab:Table 95CL} we present the upper bounds on the boost factor (assuming $\langle\sigma_{ann}\vert v\vert\rangle = 3\times 10^{-26}\;  \rm cm^{3} s^{-1}$) for several XDM-type annihilation channels.  For the XDM $e^{\pm}$, XDM $\mu^{\pm}$, XDM $\pi^{\pm}$ channels, we used the mean efficiency $f$ provided in Table 1 of \cite{Slatyer:2009yq}.  For XDM 1:1:2 the efficiency is a linear combination of the efficiencies for XDM $e^{\pm}$, XDM $\mu^{\pm}$ and XDM $\pi^{\pm}$, while for XDM 2-step to $e^{\pm}$ and 2-step to $\mu^{\pm}$, the efficiency is approximately equal to the efficiency of XDM $e^{\pm}$ and XDM $\mu^{\pm}$, respectively.

The values of $BF_{WMAP5}$ presented in the table are upper limits on the allowed BF from Sommerfeld enhancement at $z \sim 1000$.  Since the velocity dispersion of DM in both the spherical halo and dark disk are expected to be larger now than at $z \sim 1000$ (resulting in smaller boost factors now), the values in Table~\ref{tab:Table 95CL} can be considered the maximum allowed values of the boost factors today (95 $\%$ C.L.).  Current values of the annihilation cross-section would be equal to those at $z \sim 1000$ if we are in the saturated regime.

A comparison of the BF's for the fits shown in Figs.~\ref{fig:XDM_singlemodes}-\ref{fig:XDM_2step_muons} with the allowed ranges of $BF_{SH}$ and $R$ of Table~\ref{tab:Table 95CL} makes it clear that the CMB induced limits are low enough to constrain the contribution of the dark disk for the various channels.  We remind the reader that the BF's given in Figs.~\ref{fig:XDM_singlemodes}-\ref{fig:XDM_2step_muons} assume a local DM density of $\rho_{0} = 0.4 \; \rm GeV cm^{-3}$, and that, as discussed in Section~\ref{sec:defns}, the value could be different by a factor of $\sim$2.  If $\rho_{0}=0.6 \, \rm GeV cm^{-3}$, the BF's needed to fit the \textit{Fermi} or \textit{PAMELA} data would be a factor of $\sim 1/2$ of those shown.  Thus, a factor of a few discrepancy between the fitted BF and the CMB limit is allowed. 

For the XDM $\mu^{\pm}$ channel, the \textit{Fermi} data excludes neither the spherical Einasto halo nor the dark disk.  The boost factor for the fit of the dark disk to the \textit{PAMELA} data is $\sim 5$ times bigger than the allowed value.\footnote{Multiple species can both decrease the needed annihilation rates for the observed \textit{PAMELA} positron fraction by a factor of $\sim 2$ and evade the CMB limits.  See \cite{Cholis:2009va} for a more extended discussion on indirect signals from multiple species of annihilating DM.}  (Similar results hold for the XDM $e^{\pm}$ channel whose fits to the data are not shown.)  However, if the contribution to the local $e^{\pm}$ flux of the dark disk and the spherical halo are comparable at $\sim 500$ GeV, the fluxes at \textit{PAMELA} energies come mainly from the spherical halo.  Thus, even if the dark disk gives an $O(1)$ contribution to the local flux at $E\sim 10^{3}$ GeV, those channels are not yet excluded by the CMB limits, but are in some tension.

The XDM $\pi^{\pm}$ and XDM 1:1:2 channels are within CMB constraints, assuming the factor of 2 discrepancy between the fitted BF and the CMB limit mentioned earlier.  In the combination channels, less of the available energy from DM annihilation is dumped into $e^{\pm}$ at $z\sim 1000$, so there is a smaller effect on the CMB temperature and polarization power spectra.  For the XDM 2-step to $e^{\pm}$ and 2-step to $\mu^{\pm}$, the \textit{PAMELA} fits require BF's within a factor of a few of the CMB limits, thus those channels aren't excluded.  We note that regardless of the specifics of the annihilation channel (and the needed DM mass and BF), the fits to the \textit{PAMELA} and \textit{Fermi} data constrain the spectrum of DM $e^{\pm}$ to be similar in all cases, thus we expect the CMB limits to be similarly constraining in all cases.

Recently \cite{Feng:2010zp, Feng:2009hw} suggested stronger constraints on the Sommerfeld enhancement based on relic density calculations.  In \cite{Feng:2010zp} the authors suggest that for $m_{\chi} = 1$ TeV and $m_{\phi}=250$ MeV, the maximum allowed Sommerfeld enhancement is 90.  In general, they find that for $m_{\chi} > 1$ TeV and $m_{\phi}\sim 1$ GeV the allowed Sommerfeld enhancement is $\sim100$.  These values were calculated assuming a velocity dispersion of ~200 km/s, and thus can be compared to our boost factors for the spherical halo.  However, it is unclear what the results would be for the dark disk, which has a velocity dispersion that is $\sim5$ times smaller.  We note that in \cite{Feng:2010zp} they assume the Sommerfeld enhancement reaches the saturation bound at a much lower velocity dispersion than that of the dark disk or the spherical halo.  If the saturation bound occurs at velocities of $\sim 200$ km/s, or even $\sim 30$ km/s, the effect of the Sommerfeld enhancement on the relic density is much smaller, since much higher annihilation cross-sections at earlier epochs are not allowed.  Thus we still consider as a reference the constraints of \textit{WMAP}5 CMB power spectra from \cite{Slatyer:2009yq}.  We expect that data from \textit{Planck} \cite{:2006uk} will make these constraints significantly tighter.

\subsection{Photons}

The predictions for the photon signals from dark matter annihilation are highly dependent on the distribution of the photon sources, i.e. the dark matter, in the Galaxy.  Because photons travel to our detectors from their place of origin without diffusion and without loss of energy, their fluxes sample their source distribution very well.  Thus, not only is the relative density of dark matter in the disk to that in the spherical halo ($\rho_{0_{DD}}/\rho_{0_{SH}}$) important for calculating photon fluxes, but so is the relative annihilation cross-section.  Whether or not there is a Sommerfeld enhancement and whether or not it is in saturation is quite relevant for photons.  For a discussion of the range of reasonable values of the ratio of densities, see Section~\ref{sec:defns}.  There are two limiting cases: $\rho_{DD_{0}}/\rho_{SH_{0}} \sim 1$ and $\rho_{DD_{0}}/\rho_{SH_{0}} \ll 1$.

\vspace{\baselineskip}
\noindent{\textit{Microwave haze}}

Cosmic ray electrons and positrons with GeV-scale energies will synchrotron radiate through their interactions with the Galactic magnetic field, resulting in microwave frequency radiation.  The dark matter models that fit the \textit{PAMELA} and \textit{Fermi} electronic data predict a population of $e^\pm$ with the right spectrum and spatial distribution \cite{Finkbeiner:2004us, Hooper:2007kb, Cholis:2008wq}, assuming a spherical DM halo) to account for the \textit{WMAP} microwave ``haze'' found by \cite{Finkbeiner:2003im} and later confirmed by \cite{Dobler:2007wv}.  We have calculated the effect of the addition of a dark disk on the synchrotron radiation in the haze region, i.e. the average value over longitudes $-10^{\circ}$ to $+10^{\circ}$ for latitudes between $6^{\circ}$ and $15^{\circ}$ from the Galactic Plane.

Here we assume locally equal, $\rho_{DD_{0}}/\rho_{SH_{0}} \sim 1$, and spatially uniform annihilation cross-sections for the DM particles in the dark disk halo and the spherical halo.  This case trivially applies for annihilating DM models like Kaluza-Klein DM as presented in \cite{Servant:2002aq, Cheng:2002ej} and studied by \cite{Baltz:2004ie, Hooper:2007kb, Hooper:2009fj} (among others), where the Sommerfeld enhancement cannot be used to explain large boost factors.  As we have already discussed, the Sommerfeld enhancement may be at or very close to saturation, so the assumption of equal annihilation cross-sections may be valid for theories that include the Sommerfeld enhancement as well.  However, the annihilation cross-section may drop near the Galactic Center, since recent simulations in which baryons are included \cite{RomanoDiaz:2008wz, RomanoDiaz:2009yq, Pedrosa:2009rw, fabioprivate} suggest that the velocity dispersion grows toward the GC.  \cite{Pedrosa:2009rw} find that the dependence of the velocity dispersion $\sigma_{v}$ on the distance from the Galactic Center $r$ goes as $\sigma_{v} \sim r^{-1/4}$.

For the dark disk and Einasto profiles of Eqs.~\ref{eq:DD_eq} and~\ref{eq:SHM_eq}, under the assumption of equal local densities, the flux of synchrotron radiation for the spherical profile is $\sim 20$ times larger than that for the dark disk within the region of the microwave haze.  This value is nearly constant over the entire region.  When averaging over the inner 2 kpc, the Einasto profile has a number density squared value $\overline{n^{2}}$ that is $\sim 60$ times larger than that of the dark disk profile, $\overline{n_{SH}^{2}}/\overline{n_{DD}^{2}} \sim 60$, and about $\sim 5$ times larger than the mixed term (see Eq.~\ref{eq:annih_rate_Somm}), $\overline{\rho_{SH}^{2}}/\overline{2 \rho_{SH} \cdot \rho_{DD}} \sim 5$.  If we define $\rho_{SH}'$ to represent the DM density in the case where there is only a spherical halo, then averaging over the inner 2 kpc, assuming $\rho_{SH}' = \rho_{SH}$ and $\rho_{0_{SH}} \sim \rho_{0_{DD}}$, we find $(\rho_{SH} + \rho_{DD})^{2} / \rho_{SH}'^{2} \approx 5/4$. 
Since we measure the total local DM density, we need to normalize the total DM density locally to the same value, $ \rho_{0_{SH}} + \rho_{0_{DD}} = \rho_{0_{SH}}'$. For $\rho_{SH_{0}} \sim \rho_{DD_{0}}$ we get, $\rho_{0_{SH}}' = 2\rho_{0_{SH}}$ $\rightarrow$ $\rho_{SH}' = 2\rho_{SH}$. Thus after averaging over the inner 2 kpc $(\rho_{SH} + \rho_{DD})^{2} / \rho_{SH}'^{2} \approx 1/3$.  
If we also normalize the BF to fit the local $e^{\pm}$ fluxes and include the effects of propagation of the highest energy $e^{\pm}$, then in non-Sommerfeld cases and saturated Sommerfeld cases we still need approximately the same BF to explain the \textit{WMAP} haze, even including a dark disk.  However, if the Sommerfeld enhancement for the disk is significantly greater than for the spherical halo (and still within constraints), then the synchrotron radiation from the $e^{\pm}$ of DM origin can be decreased by a factor of $\sim 2$. \footnote{If for the dark disk the annihilation cross-section is close to a resonance, then the effect on the synchrotron radiation from the $e^{\pm}$ of DM origin is even more prominent.}

Common variations of dark disk profile of Eq.~\ref{eq:DD_eq} include \cite{Read:2008fh}:
\begin{equation}\label{eq:other_DD_profiles}
\rho(R,z) \sim \exp\left[-\frac{R}{R_{1/2}}\right] \exp\left[-\Bigr(\frac{z}{z_{1/2}}\Bigr)^{2}\right]\;\;\mbox{and}\;\;\; \nonumber
\rho(R,z) \sim \exp\left[-\frac{R}{R_{1/2}}\right] \mbox{sech}^{2}\left[-\frac{z}{z_{1/2}}\right].
\end{equation}
Taking $z_{1/2}=1.5 \; \rm kpc$ and $R_{1/2}=11.7 \; \rm kpc$, we calculate synchrotron fluxes similar to those for our reference dark disk. Thus, the above arguments apply for these variations as well. \cite{Kalberla:2007sr} has suggested a dark disk with $z_{1/2}=2.8 \; \rm kpc$ and $R_{1/2}=12.6 \; \rm kpc$, using the HI gas distribution as an alternative tracer of the gravitational field.  Using these values of the scale lengths in our profile of Eq.~\ref{eq:DD_eq} results in an increase in the synchrotron flux from the dark disk of 50\% over the flux calculated using the values of $z_{1/2}=1.5 \; \rm kpc$ and $R_{1/2}=11.7 \; \rm kpc$. Thus, the specifics of the profile of the dark disk have a small influence on our conclusions.  We note that since the \textit{Fermi} ($\gamma$-ray) haze \cite{Dobler:2009xz} is the $\gamma$-ray counterpart of the \textit{WMAP} haze, the previous discussion applies to the \textit{Fermi} haze as well.

\vspace{\baselineskip}
\noindent{\textit{Diffuse $\gamma$-rays}}

In addition to producing synchrotron radiation of microwave wavelengths through interactions with the Galactic magnetic field, the population of $e^\pm$ from dark matter annihilation can produce gamma rays through inverse Compton scattering (ICS) off of the optical, infrared, and CMB photons in the interstellar medium.  Another source of $\gamma$'s from DM annihilation is prompt emission, gammas produced in the decay chains of the annihilation products.

\begin{figure}[t!]
\centering
\begin{tabular}{cc}
\epsfig{figure=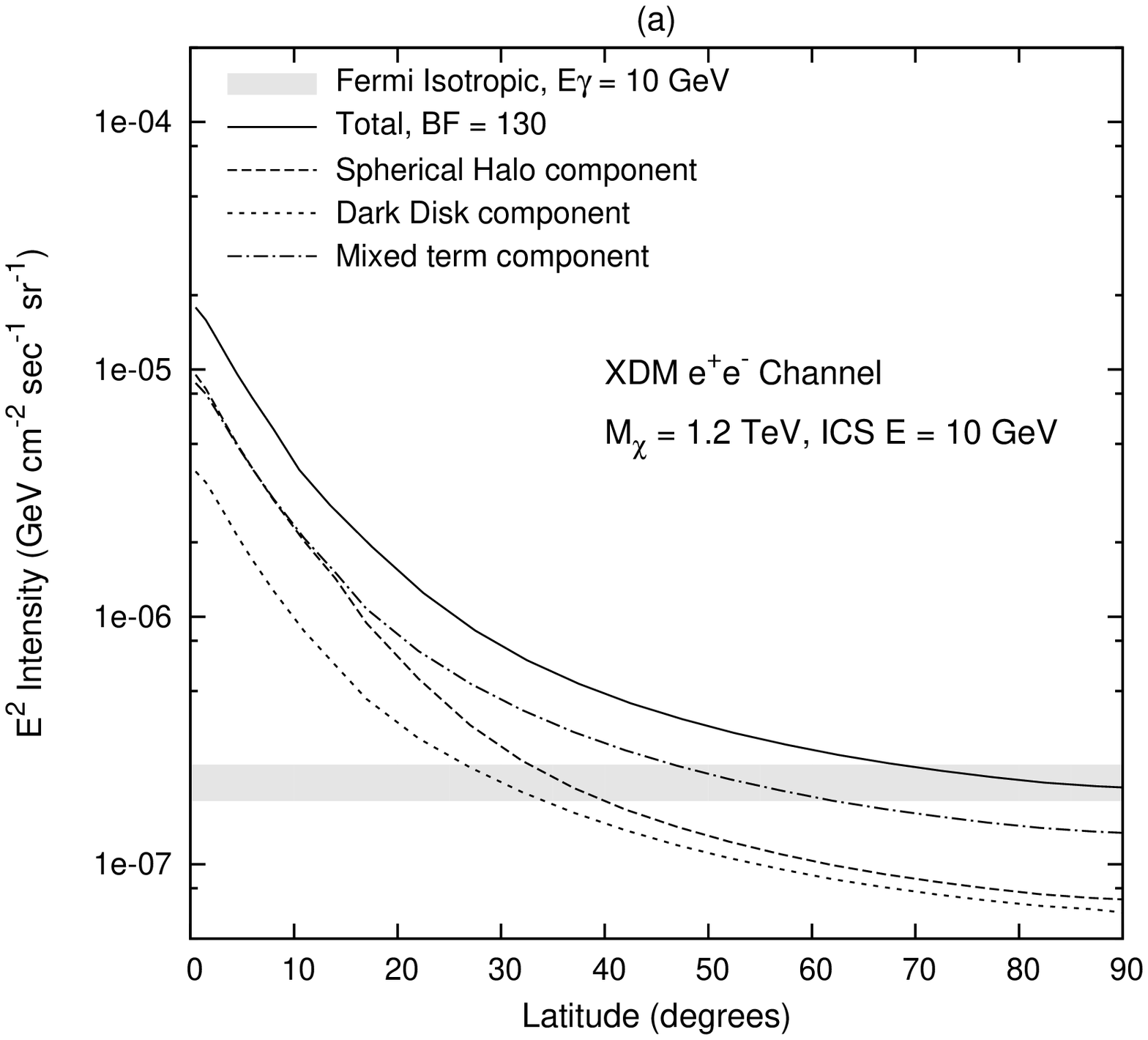,width=3.0in} &
\epsfig{figure=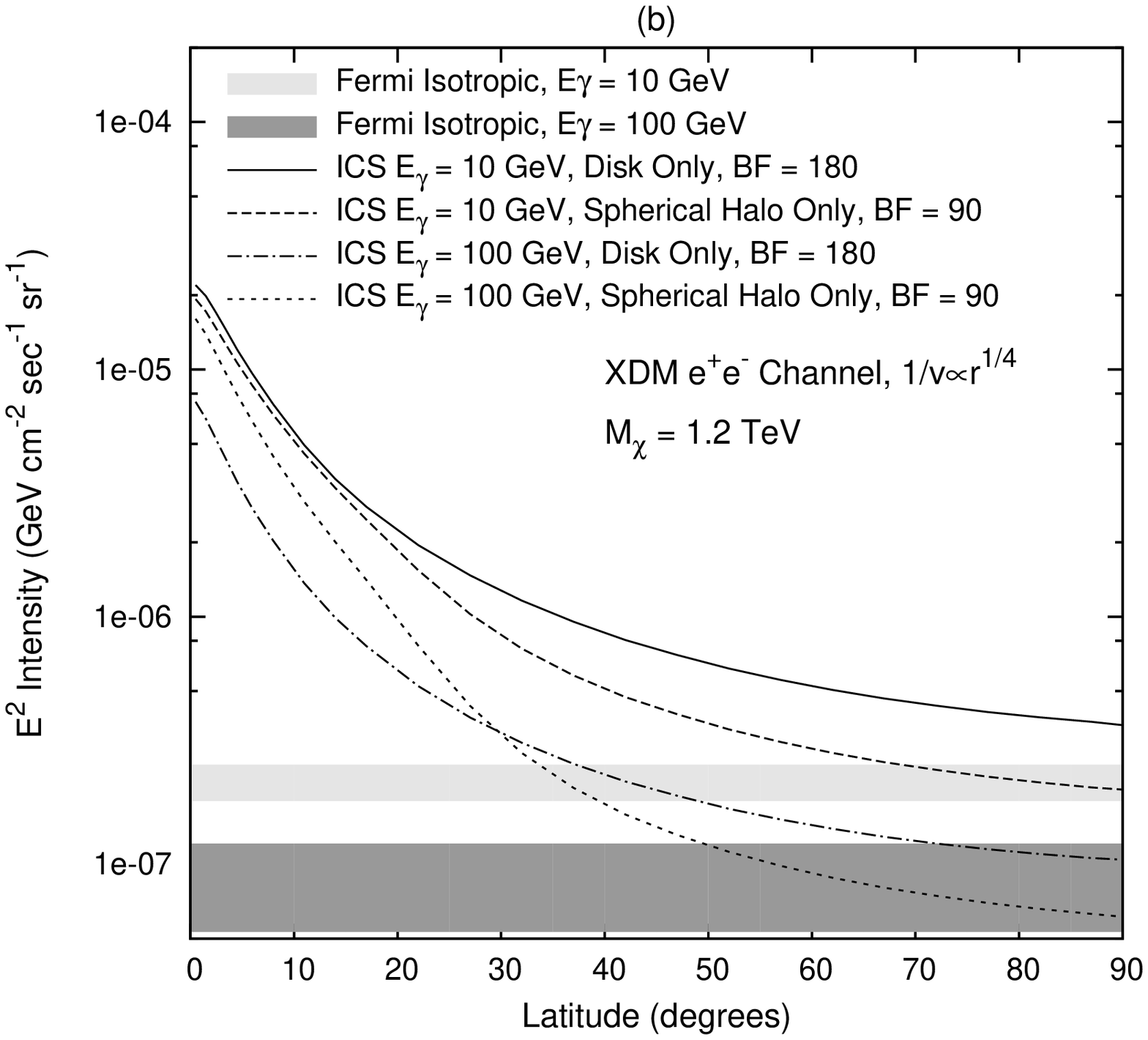,width=3.0in} 
\end{tabular}
\caption{$\gamma$-ray flux from ICS of DM $e^{\pm}$ off the ISRF as a function of galactic latitude for the XDM $e^{\pm}$ channel with $m_{\chi}=1.2$ TeV.  Left: $E_{\gamma}=10$ GeV, saturated Sommerfeld enhancement.  Right: Sommerfeld enhancement with annihilation in the dark disk only and in the spherical DM halo only for $E_{\gamma}=$ 10 and 100 GeV.  The isotropic diffuse $\gamma$-ray flux measured by \textit{Fermi} \cite{Abdo:2010nz} at 10 GeV and 100 GeV(extrapolated) is shown within 2$\sigma$.}
\label{fig:Latitudinal_profiles}
\end{figure}

In Fig.~\ref{fig:Latitudinal_profiles}a we show the flux of 10 GeV \footnote{$E_{\gamma}\sim 100$ GeV is also relevant for probing DM, but \textit{Fermi} statistics are very low and CR contamination is more significant around 100 GeV.} ICS $\gamma$-rays for the XDM $e^\pm$ channel, averaged over longitudes $| l | < 5^{\circ}$, as a function of latitude $b$.  Here we assume equal local DM densities in the spherical and disk halos, and normalize the local $e^{+}+e^{-}$ flux to fit the \textit{Fermi} data.
We also indicate in the plot the isotropic diffuse $\gamma$-ray flux at 10 GeV as measured by \textit{Fermi} \cite{Abdo:2010nz} with 2$\sigma$ errorbars.\footnote{The flux is based on the data shown in \cite{Abdo:2010dk} and \cite{Collaboration:2010gq}.}  As recently suggested by \cite{Collaboration:2010gq}, about $\sim 20\%$ of the isotropic diffuse $\gamma$-ray flux for $E>0.1$ GeV is attributed to unresolved extragalactic sources, mainly blazars, with a power-law of $E^{-2.2}$.  (See also \cite{Abdo:2009iq}.)  Thus, it is possible that some of the isotropic diffuse $\gamma$-ray flux is coming from dark matter annihilation or decay.  For the scenario shown in Fig.~\ref{fig:Latitudinal_profiles}a, the galactic diffuse $\gamma$-ray flux for $| b | > 70^{\circ}$ is very similar in magnitude to the isotropic diffuse flux.  It is thus evident that constraints can be put on the annihilation rate of DM models from diffuse $\gamma$-rays at high latitude.  On a related note, \cite{Zavala:2009zr} and \cite{Abdo:2010dk} have suggested that dark matter substructure at $z>0$ can contribute an important fraction to the isotropic $\gamma$-ray flux.   \cite{Abdo:2010dk, Abazajian:2010sq, Hutsi:2010ai} have shown that calculations of the extragalactic $\gamma$-ray flux from cosmological DM can also put constraints on the annihilation cross-section of various DM models.

Including a dark disk increases the galactic diffuse $\gamma$-ray flux at high latitudes ($| l | > 60^{\circ}$) by 1/2 relative to the flux at low latitudes (near the GC), in cases where the main production mechanism of 10 GeV $\gamma$-rays from dark matter annihilation is ICS of $e^{\pm}$ off the Interstellar Radiation Field (ISRF).  This is readily seen in Fig.~\ref{fig:Latitudinal_profiles}a by comparing the flux from the Spherical Halo to the total flux; the flux from the SH falls off more quickly with latitude.  Therefore, a significant disk component of dark matter could result in stronger constraints coming from the isotropic diffuse $\gamma$-ray flux than from the diffuse flux at the Galactic Center.  For DM models producing a large fraction their total gamma rays at $E_{\gamma} =10$ GeV coming from prompt emission, the relative increase of high latitude flux to low latitude flux is smaller.  We therefore conclude that a dark disk can't be a significant issue in the search for $\gamma$-ray anisotropies due to DM clumps \cite{Diemand:2006ik, Kuhlen:2008aw, SiegalGaskins:2008ge, Bovy:2009zs, Kuhlen:2009jv, Kistler:2009xf, SiegalGaskins:2009ux, Hensley:2009gh} at high latitudes.

We now consider a situation in which the Sommerfeld enhancement is not in saturation, so that the annihilation cross-sections in the disk and spherical halo may be quite different.  We assume $\rho_{0_{DD}}/\rho_{0_{SH}} \sim 1$.  In Fig.~\ref{fig:Latitudinal_profiles}b, we show the $\gamma$-ray flux from ICS of DM $e^{\pm}$ off the ISRF as a function of galactic latitude for the two extreme cases, one in which the annihilation in the spherical halo is dominant, i.e. we ignore any dark disk contribution, and one in which the annihilation in the dark disk is dominant, i.e. we ignore any spherical halo contribution.  We show results for the XDM $e^{\pm}$ annihilation channel with $M_{\chi}=1.2$ TeV, and also include the \textit{Fermi} isotropic flux at $E_{\gamma}=10$ GeV and 100 GeV.  We normalize the annihilation rates in both cases to give agreement with the local $e^{+}e^{-}$ CR spectra.  The diffuse $\gamma$-ray flux at high latitudes is larger by a factor of 2(4) at $E_{\gamma}=10 (100)$ GeV in the disk-dominated scenario than in the spherical halo-dominated scenario.  Thus, even if the Sommerfeld enhancement is currently not saturated, the existence of a dark disk with a velocity dispersion of $\sim 30$ km/s does not significantly increase the high latitude diffuse $\gamma$-ray flux.  Clearly, Fig.~\ref{fig:Latitudinal_profiles}b indicates that the annihilation rate in the dark disk is highly constrained by the isotropic diffuse flux at $E_{\gamma}=10$ GeV.

Since the velocity dispersion increases with decreasing galactocentric distance, annihilation cross-sections toward the center of the Galaxy will be smaller in models of DM with a Sommerfeld enhancement, provided the enhancement is not saturated.  So, the constraints on the BF needed to fit the local CR $e^{\pm}$ fluxes coming from the $\gamma$-rays measured by HESS in the Galactic Ridge (GR) region and inner $0.1^{\circ}$ of the Galactic Center \cite{Aharonian:2006au, Aharonian:2004wa, Aharonian:2006wh, Aharonian:2009zk} are lifted, as was shown in \cite{Cholis:2009va}.  While the existence of a dark disk boosts the DM signals of local origin, the $e^{\pm}$ signals, while at the same time not dramatically increasing the $\gamma$ signals from the GC, the real gain to be had in Sommerfeld-enhanced models comes from the effects of the increase in velocity dispersion toward the GC.  For DM models without Sommerfeld enhancement, the constraints placed on the local annihilation rates by the gamma ray fluxes in the GR and GC regions \cite{Bell:2008vx, Bertone:2008xr, Cirelli:2009vg, Crocker:2010gy} are similar with and without a dark disk component.

\section{Conclusions}\label{sec:Concusions}

A self-consistent explanation of the \textit{PAMELA} and \textit{Fermi} signals as coming from dark matter annihilation requires a local annihilation rate in the spherical halo that is 2 to 3 orders of magnitude larger than that expected for a thermal WIMP.  The large boost factor could arise from substructure in the halo (a local inhomogeneity), from the Sommerfeld enhancement, or from a Breit-Wigner resonance, or it could be an indication of a non-thermal WIMP.  We consider effects of a dark disk component with a local density of $0.4 \;\rm GeV cm^{-3}$.  The lower velocity dispersion of dark matter particles in the dark disk can naturally lead to higher annihilation rates through the Sommerfeld enhancement, which scales as $1/v_{rel}$.  We show that if the dominant dark matter contribution to the local high energy $e^\pm$ flux comes from a dark disk, rather than the spherical halo, we can get good fits to the \textit{Fermi} and \textit{PAMELA} with a boost factor that is almost an order of magnitude smaller than what is needed for the fits to the data when only the spherical halo profile is taken into account.  Thus, the Sommerfeld enhancement for the spherical halo with a velocity dispersion of $v_{rel} \sim 200$ km/s need not take the extremely large values previously thought.
We stress that many of our conclusions would hold to varying degrees for all flattened halo profiles, even those with kinematics similar to the spherical halo, since all of the cosmic ray dark matter signals are highly dependent on the spatial distribution of the dark matter.

We show that the locally observed spectra of $e^{+}e^{-}$ are harder for a dark disk profile than for a spherical profile. This can improve the fits to the \textit{Fermi} data for those annihilation modes with softer $e^{+}e^{-}$ spectra.  Some tension arises as a result of adding a disk component to the dark matter density, because the discrepancy between the boost factors needed to fit the \textit{PAMELA} positron fraction and to fit the \textit{Fermi} data increases with the addition of a disk component. The spherical halo gives differences of factors of $\sim 2-3$, while the dark disk gives differences of factors of up to $\sim 9$ (see Figs.~\ref{fig:XDM_singlemodes}-\ref{fig:WW_positrons}).  One explanation is that the spherical halo contribution to the $e^{+}e^{-}$ flux is dominant at the lower energies measured by \textit{PAMELA}, while for energies greater than 300 GeV, the \textit{Fermi} range, the dark disk contribution to the $e^{+}e^{-}$ flux becomes comparable. That, however, depends on the assumptions about the contribution to the local DM density from the dark disk and the relative annihilation cross-sections in the dark disk and the spherical halo.  Furthermore, the possibility of multi-component DM or nearby pulsars contributing significantly to the local spectra decreases the significance of a difference in the boost factors necessary to fit \textit{PAMELA} vs \textit{Fermi}.

We find that for hadronic annihilation channels, if the annihilation in the disk is the dominant contribution to the local cosmic ray fluxes of dark matter origin (and if the disk is thicker than the diffusion zone), the data allows for an annihilation rate that is larger by a factor of $\sim 6$ than the rate for a spherical profile only.  Even so, the \textit{PAMELA} antiproton data constrains the annihilation products to be mainly leptonic for masses of $m_{\chi}\sim$ 1.0 TeV.

We investigate possible constraints on the dark disk contribution to the $e^\pm$ fluxes using the \textit{Fermi} $e^{+}+e^{-}$ data and CMB data.  The \textit{Fermi} data is consistent with a scenario in which the dark disk is the dominant contributor to the local $e^{+}e^{-}$ flux for the XDM annihilation channels we studied.  With the data currently available, it is impossible to determine whether the local $e^{+}e^{-}$ flux is produced dominantly in one halo component or approximately equally in both the disk and the spherical halo.  In fact, due to the many possible dark matter masses and annihilation modes in the production of the local $e^{+}e^{-}$, it will be very hard to tell the difference even with better data.  The constraints on the DM annihilation rate coming from the \textit{WMAP5} measurements of the temperature and polarization power spectra of the CMB do not exclude any of the XDM annihilation channels studied in a disk-dominated scenario.  However, because the dark disk produces fewer electrons and positrons at \textit{PAMELA} energies than the spherical halo, there is more tension for the disk than for the spherical halo between the BF needed to fit the \textit{PAMELA} data and the CMB constraint. 

We find that the contribution from the dark disk can be up to $\sim 1/2$ of the local $e^{\pm}$ flux of dark matter origin for $E_{e^\pm} > 500$ GeV.  Additionally, we find that including a dark disk has a minimal effect on the flux of synchrotron radiation from dark matter $e^\pm$ in the microwave haze region.  This result is valid even for disks with relatively large scale heights ($\sim 3$ kpc) perpendicular to the plane.  Moreover, including a dark disk of scale height $\sim 1$ kpc does not affect the diffuse $\gamma$-ray flux at high latitudes by more than $\mathcal{O}(1)$.  If the Sommerfeld enhancement is not saturated, so that the annihilation rate in the disk is much greater than the annihilation rate in the spherical halo, the contribution to the gamma ray flux from the disk at 10 GeV exceeds the value of the isotropic $\gamma$-ray flux as measured by \textit{Fermi} by a factor of $\sim 2$ at the highest latitudes where the isotropic component is the dominant component.  Thus, the annihilation rate in the dark disk is highly constrained by the isotropic diffuse flux at $E_{\gamma}=10$ GeV.

\vskip 0.2in
\noindent {\bf Acknowledgments}

IC and LG are supported by DOE OJI grant \# DE-FG02-06ER41417.  IC is also supported by the Mark Leslie Graduate Assistantship.  The authors would like to thank Jo Bovy, Douglas Finkbeiner, Michael Kuhlen, Dmitry Malyshev, Rob Morris and Jennifer Siegal-Gaskins for the valuable discussions they provided.  We are especially indebted to Neal Weiner for initiating this project, and for his support and invaluable discussions during the evolution of this work. 
\appendix
\begin{center}
\bf{Appendix A: 2-Step} $e^{+}e^{-}$ \bf{spectra} \label{sec:App_spectra}
\end{center}

The probability distribution function of injected electrons created by the annihilation channel $\chi\chi \rightarrow \Phi\Phi \rightarrow 4\phi \rightarrow (4 \mu^\pm \rightarrow)\; 4 e^\pm$ can be calculated analytically or numerically. In the frame of $\phi$, it is the convolution of the spectra of electrons and positrons in that frame with the spectrum of $\phi$ as seen from the frame in which annihilation takes place.  For simplicity in our analytic calculations, we consider $\Phi$ and $\phi$ to be scalars so that the decay products have an isotropic distribution in the rest frame of the parent particle.  The difference in the $e^{\pm}$ spectra resulting from a vector mediator is a spectral hardening of $\mathcal{O}(0.1)$ at the highest $\sim 10\%$ of the energies.  Since the deviations appear only at the highest energies where experiments have their largest errorbars, the spectra for the scalar mediator case can be considered a good approximation to the spectra for the vector mediator case.  In both cascades described, the multiplicity of electrons (positrons) produced per annihilation is 4. Our results agree with those of \cite{Mardon:2009rc}. 

For the "direct" case, $\phi \rightarrow e^\pm$, the distribution in the approximation $m_{\chi}\gg m_{\Phi} \gg m_{\phi} \gg m_{e}$ is given by:

\begin{equation}	
f\left( E \right)=\frac{1}{m_{\chi}}\ln\Bigl(\frac{m_{\chi}}{E}\Bigr) \;\Theta \Bigl(m_{\chi}-E\Bigr).
\label{eq:2_step_direct}
\end{equation} 
For the case of $\phi$$\rightarrow$ $\mu^{+}\mu^{-}$ and consequently muon decay, the spectrum's given by\footnote{assuming $m_{e}=0$ and $\gamma_{1}>10$ which are valid for the cases at hand.}:
\begin{eqnarray*}	
f(E)&=&\Bigl[\frac{5}{6}\frac{m_{\mu}}{W\; m_{\phi}}\Bigl(\frac{1}{2}\frac{1}{\gamma_{2}-\gamma_{1}}\ln^{2}\Bigl(\frac{\gamma_{2}}{\gamma_{1}}\Bigr) - \frac{1}{\gamma_{2}-\gamma_{1}}\ln\Bigl(\frac{2\gamma_{1}W\; {m_{\phi}}}{E\; m_{\mu}}\Bigr)\cdot\Bigl(\frac{\gamma_{2}}{\gamma_{1}}-1-\ln(\frac{\gamma_{2}}{\gamma_{1}}\Bigr)\\
&+&\frac{1}{\gamma_{1}}\ln\Bigl(\frac{2\gamma_{1}W\; {m_{\phi}}}{E m_{\mu}}\Bigr)\Bigr)
+\frac{19}{36}\frac{m_{\mu}}{W\; m_{\phi}}\Bigl(\frac{1}{\gamma_{2}-\gamma_{1}}\cdot\Bigl(\frac{\gamma_{2}}{\gamma_{1}}-1-\ln\Bigl(\frac{\gamma_{2}}{\gamma_{1}}\Bigr)\Bigr)-\frac{1}{\gamma_{1}}\Bigr)\\
&-&\frac{3}{32}\; \Bigl(\frac{m_{\mu}}{W\; m_{\phi}}\Bigr)^{3}E^{2}\Bigl(\frac{1}{\gamma_{2}-\gamma_{1}}\Bigl(\frac{1}{\gamma_{2}^{2}}-\frac{3}{\gamma_{1}^{2}}+\frac{2 \gamma_{2}}{\gamma_{1}^{3}}\Bigr)-\frac{2}{\gamma_{1}^{3}}\Bigr)\\
&+&\Bigl(\frac{m_{\mu}}{W\; m_{\phi}}\Bigr)^{4}E^{3}\Bigl(\frac{1}{\gamma_{2}-\gamma_{1}}\Bigl(\frac{1}{108\gamma_{2}^{3}}-\frac{4}{108\gamma_{1}^{3}}+\frac{4\gamma_{2}}{144\gamma_{1}^{4}}\Bigr)-\frac{1}{36\gamma_{1}^{4}}\Bigr)\Bigr]\;\Theta \Bigl(2W\gamma_{1}\frac{m_{\phi}}{m_{\mu}}-E\Bigr)\\
&+&\Bigl[\frac{m_{\mu}}{W\; m_{\phi}}\Bigl(\frac{65}{216}\frac{1}{\gamma_{2}-\gamma_{1}}-\frac{19}{36}\frac{1}{\gamma_{2}-\gamma_{1}}\ln\Bigl(\frac{2\gamma_{2}W\cdot{m_{\phi}}}{E\; m_{\mu}}\Bigr)\\
&+&\frac{5}{12}\frac{1}{\gamma_{2}-\gamma_{1}}\ln^{2}\Bigl(\frac{2\gamma_{2}W\; {m_{\phi}}}{E\; m_{\mu}}\Bigr)\Bigr)-\frac{3}{32}\Bigl(\frac{m_{\mu}}{W\; m_{\phi}}\Bigr)^{3}E^{2}\frac{1}{(\gamma_{2}-\gamma_{1})\gamma_{2}^{2}}\\
&+&\frac{1}{108}\Bigl(\frac{m_{\mu}}{W\; m_{\phi}}\Bigr)^{4}E^{3}\frac{1}{(\gamma_{2}-\gamma_{1})\gamma_{2}^{3}}\Bigr]\;\Theta \Bigl(E-2W\gamma_{1}\frac{m_{\phi}}{m_{\mu}}\Bigr)\;\Theta \Bigl(2W\gamma_{2}\frac{m_{\phi}}{m_{\mu}}-E\Bigr).\label{eq:2_step_muon}
\end{eqnarray*} 
where
\[W = \frac{m_{\mu}}{2},\; \gamma_{1}=\frac{m_{\chi}}{2m_{\phi}}\Bigl(1 - \sqrt{1-\Bigl(\frac{2m_{\phi}}{m_{\Phi}}\Bigr)^{2}}\sqrt{1-\Bigl(\frac{m_{\Phi}}{m_{\chi}}\Bigr)^{2}}\Bigr) \mbox{, and} \;\gamma_{2}=\frac{m_{\chi}}{m_{\phi}} -\gamma_{1} \;.\]

\bibliographystyle{plain}
\bibliography{darkdisk}

\end{document}